# On the Accuracy with which the Lower Boundary Conditions can be Determined in Numerical Models of the Atmosphere

Gerhard Kramm[1] and Ralph Dlugi[2]

[1]University of Alaska Fairbanks, Geophysical Institute,
903 Koyukuk Drive, P.O. Box 757320
Fairbanks, AK 99775-7320, USA

[2]Arbeitsgruppe Atmosphärische Prozesse (AGAP),
Gernotstraße, D-80804 Munich, Germany

**Abstract**

Since the prediction of climate is mainly considered as a prediction of second kind, it is indispensable to assess the accuracy with which these boundary conditions can be determined so that we can find a reasonable answer, whether climate is predictable with a sufficient degree of accuracy or not. Therefore, our contribution is mainly focused on the accuracy with which the fluxes of sensible heat and water vapor, required for predicting the lower boundary conditions in numerical models of the atmosphere using the coupled set of energy and water flux balance equations for the Earth's surface, can be determined. The parameterization schemes for the interfacial sublayer in the immediate vicinity of the Earth's surface and for the fully turbulent layer above presented and discussed here document that an appreciable degree of uncertainty exists. It is shown that the great uncertainty inherent in the universal functions of the Monin-Obukhov similarity laws on which the parameterization schemes for the fully turbulent atmospheric surface layer are based is reflected by the considerable scatter in the results of sophisticated field campaigns. This uncertainty affects also the results for the gradient-Richardson number, the turbulent Prandtl number, $\mathrm{Pr}_t$, and the turbulent Schmidt number, $\mathrm{Sc}_{t,q}$, (and the turbulent Lewis-Semenov number, $\mathrm{LS}_{t,q}$) for water vapor used in such parameterization schemes. It is argued that the inherent uncertainty prevents that climate is predictable with a sufficient degree of accuracy.

## 1. Introduction

It is well known that numerical predictions by atmospheric models need boundary conditions. Such boundary conditions may be either Neumann-type or Dirichlet-type ones.

Boundary conditions also play a prominent role in the distinction between weather predictions und climate predictions (meanwhile called climate projections). In his contribution "Is climate predictable" Klaus Hasselmann (2002), for instance, wrote: "*The meteorologist Edward Lorenz (1975), one of the founders of chaos theory, distinguished between two kinds of prediction. Predictions of first kind concern the time-dependent evolution of a system as a function of the initial conditions with fixed boundary conditions. Predictions of the second kind concern the response of a system to changes in the boundary conditions, with fixed initial conditions. Weather forecasting is clearly a prediction problem of the first kind, while the prediction of the climate change due to human influences is normally regarded as a problem of* *the second kind.*" He eventually argued that the computation of anthropogenic climate change represents a mixed prediction problem of the first and the second kind. The same, of course, is true in the case of weather predictions. From this point of view, the distinction between weather predictions und climate projection seems to be rather arbitrary. According to Hasselmann (2002) another aspect has to be considered, too. Let us quote him: "*By definition, weather and climate prediction are mutually exclusive. As a chaotic system, the atmosphere is unstable with respect to small perturbations of the initial conditions. This sets a natural limit to the predictability of detailed weather properties. Theoretically, the prediction limit is of the order of 20 days (Lorenz, 1967). … Climate, in contrast, is defined in terms of the statistics of weather. The data ensemble used to form the statistics is required, in modern definition of climate (cf. GARP, 1975), to cover a period at least as the theoretical limit of weather prediction. Climate prediction is therefore concerned with slow changes in the statistical properties of weather on time scales of 20 days and longer. In other words, climate prediction begins where weather prediction ends.*" In a footnote he further stated: "*Prior to the modern view of climate as a dynamic rather than a static system, climate was defined in terms of 30 yr averages. This had the conceptual disadvantage of introducing a spectral gap between weather and climate prediction.*" This means that this spectral gap between weather and climate predictions, i.e., two forms of applications of atmospheric modeling, served to redefine the statistical basis of climate from a set of about 11,000 daily weather states observed at each location of a network to a set of 20 or somewhat more ones. Note that in practice we are glad to predict the weather for the next 5 or 6 days with a sufficient degree of accuracy. This strong deviation from the theoretical limit is not only affected by imperfect balance equations, but also by imperfect initial and boundary conditions.

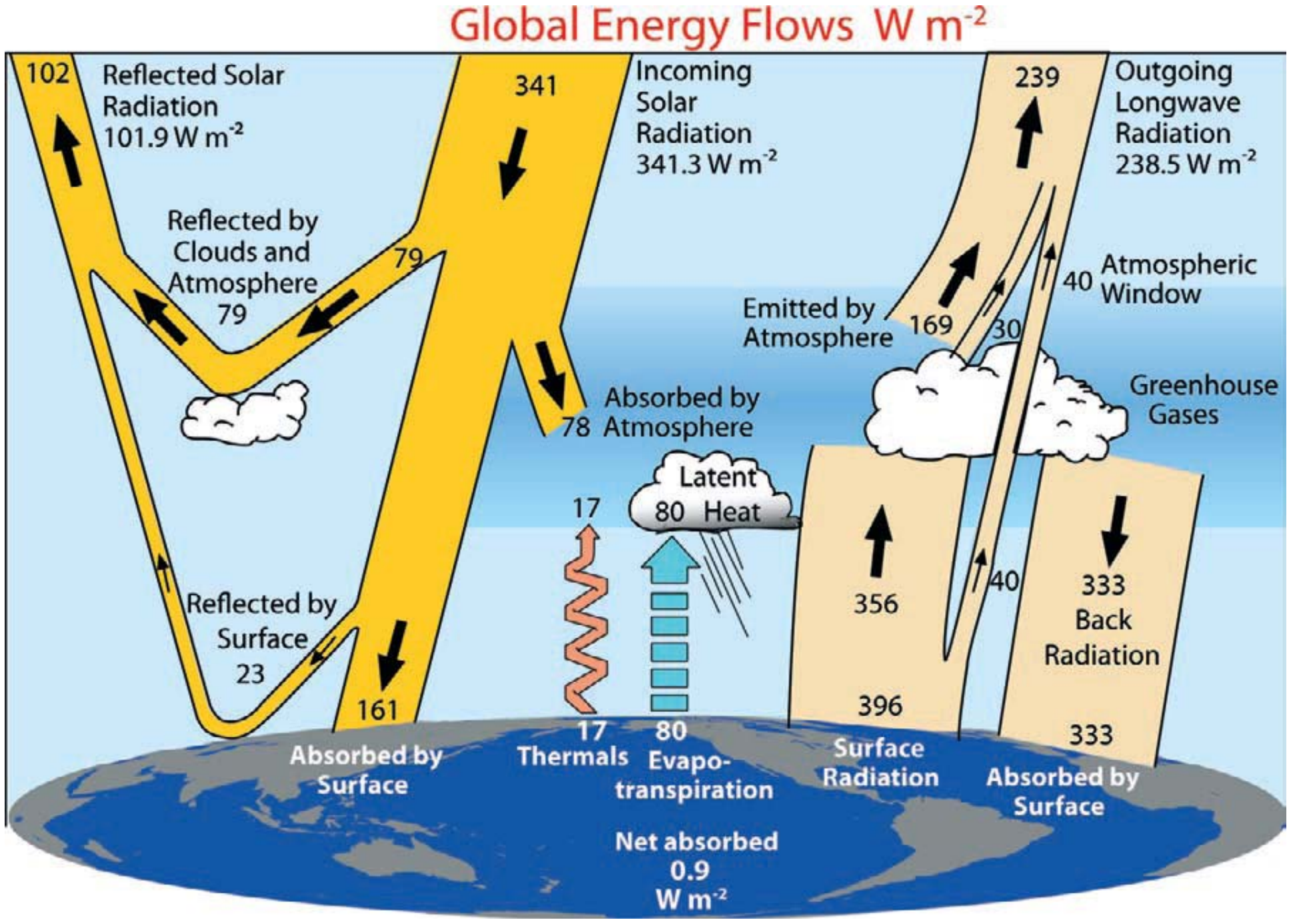

**Figure 1:** The global annual mean Earth's energy budget for the Mar 2000 to May 2004 period (W $m^{-2}$). The broad arrows indicate the schematic flow of energy in proportion to their importance (adopted from Trenberth et al., 2009).

Since there is no conservation equation of the radiation intensities, solving the radiative transfer equation (RTE) in a reliable manner needs appropriate boundary conditions. At the top of the atmosphere (TOA) a radiative equilibrium is usually postulated (see Figure 1). Such equilibrium, however, can only exist when the incoming solar radiation minus the portion reflected into the space and the outgoing infrared radiation emitted by the Earth and the atmosphere are averaged over the entire globe for long periods. The solar irradiance at the TOA, for instance, is ranging from $1420\ \mathrm{W\ m^{-2}}$ at Perihelion and $1330\ \mathrm{W\ m^{-2}}$ at Aphelion during the annual cycle of the Earth's revolution around the sun (e.g., Iqbal, 1983; Liou, 2002). As shown in Figure 2, the maximum of the spectrum of solar radiation related to the TOA by a mean distance between the Sun's center and the Earth's orbit of about $149.6 \cdot 10^6\ \mathrm{km}$ also differs from that of the terrestrial radiation for a surface temperature of 288 K by nearly two orders of magnitude. Additionally, the latter is nearly four times higher than that for a typical temperature of the tropopause of about 223 K. Obviously, we have to consider diurnal variations and seasonal variation. The latter is related to the oblique angle of the Earth's axis. On long-term scales we also must pay attention to the Milanković (1941) climate cycles related to the change of the eccentricity (400 and 100 kyear) and the oblique angle (axial tilting: 41 kyear) and the axial precession (23 kyear) (see also Berger, 1988).

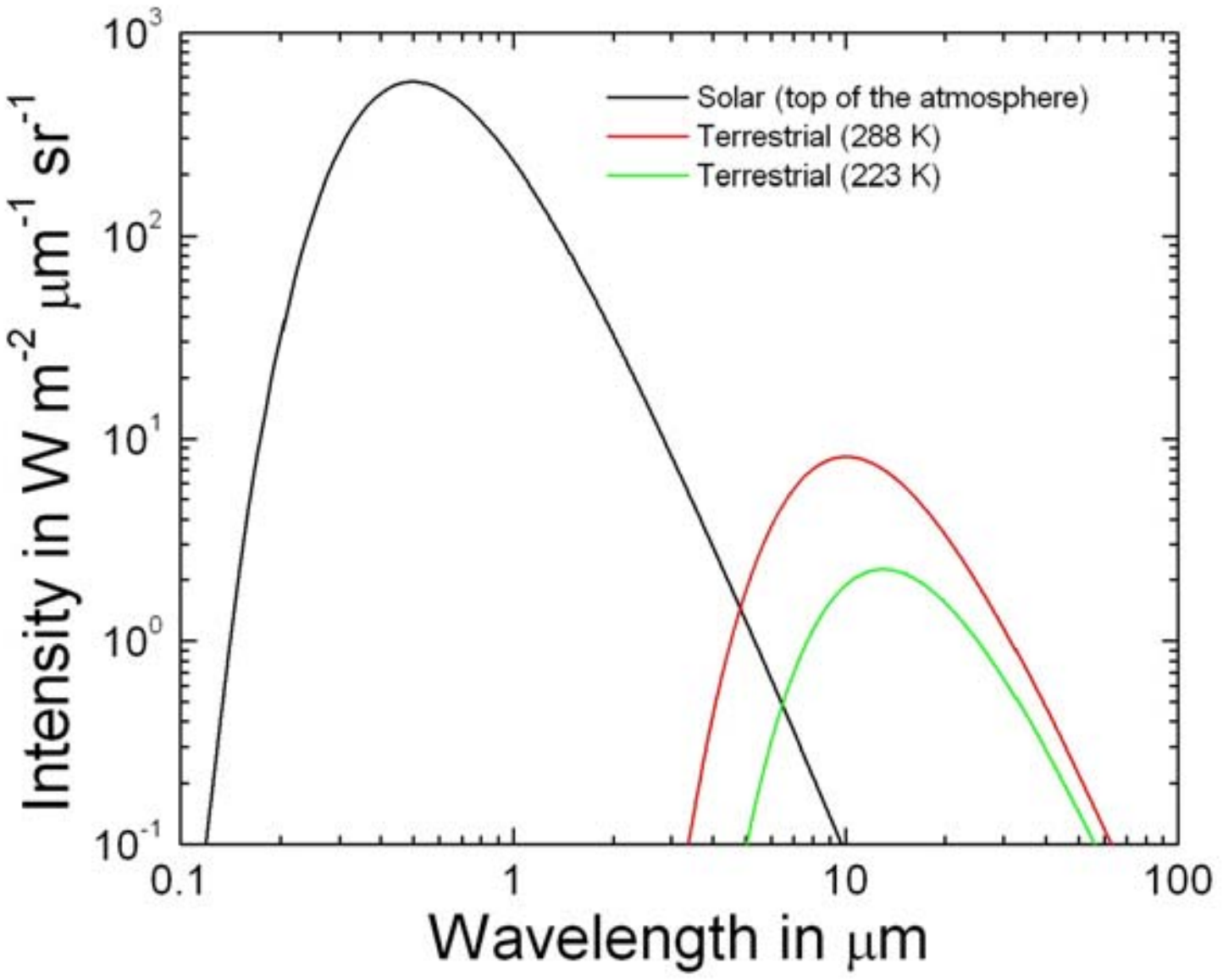

**Figure 2:** Planck functions for the solar radiation at the top of the atmosphere and the terrestrial radiation for two different temperatures.

As illustrated in Figure 1, a radiative equilibrium does not generally exist at the Earth's surface even though the so-called climate feedback equation (e.g., Schneider and Mass, 1975; Hansen et al., 1984; Dickinson, 1985; Manabe and Stouffer, 2007),

$$\mathrm{R}\,\frac{\partial \mathrm{T_E}}{\partial t} = \mathrm{Q_T} - \lambda\ \mathrm{T_E} \tag{1.1}$$

with

$$Q_T = \left(1 - \alpha_E\right) \frac{S}{4} - a + b\, T_{ref} \qquad (1.2)$$

and $\lambda = b$, is based on it. Here, $R$ is the thermal inertia coefficient, $T_E$ is the Earth's surface temperature considered as uniform, $Q_T$ is the so-called thermal forcing, $\alpha_E$ is the planetary shortwave albedo, $S$ is the solar constant, $T_{ref} = 273$ K is a reference temperature, $\lambda$ is the feedback parameter, and $a$ and $b$ are empirical constants related to Manabe and Wetherald (1967) and Budyko (1969, 1977). If $a = 211.2\ \mathrm{W\ m^{-2}}$ and $b = 1.55\ \mathrm{W\ m^{-2}\ K^{-1}}$ are chosen as recommended by Kiehl (1992), the steady-state solution of Eq. (1.1) reads $T_s(\infty) = 291.2$ K. However, if the fluxes of sensible and latent heat $H$ and $E$, as well as the absorption of solar radiation by atmospheric constituents, $A_a\, S$, are included formula (1.2) will become

$$Q_T = \left(1 - \alpha_E - A_a\right) \frac{S}{4} - H - E - a + b\, T_{ref} \quad . \qquad (1.3)$$

Recent estimates for the fluxes of sensible and latent heat result in $H \cong 17\ \mathrm{W\ m^{-2}}$ and $E \cong 80\ \mathrm{W\ m^{-2}}$ (see Trenberth et al., 2009, and Figure 1). Taking $\alpha_E = 0.30$ and $A_a = 0.23$ into account, the solar radiation absorbed by the Earth's skin results in $\left(1 - \alpha_E - A_a\right) S/4 = 160.6\ \mathrm{W\ m^{-2}}$ (see Trenberth et al., 2009, and Figure 1). If formula (1.3) is considered, the steady-state solution of Eq. (1.1) is given by $T_s(\infty) = 177.9$ K (Kramm and Dlugi, 2009). This temperature is much lower than the temperature of the planetary radiative equilibrium of about $T_e \approx 255$ K for the Earth's surface in the absence of the atmosphere. Recently, Pielke et al. (2007) thoroughly discussed what a global average surface temperature does really mean. The authors, for instance, stated that, as a climate metric to diagnose climate system heat changes (i.e., ‘‘global warming’’), the surface temperature trend, especially if it includes the trend in nighttime temperature, is not the most suitable climate metric. The result of $T_s(\infty) = 177.9$ K provided by the climate feedback equation in the case of realistic boundary conditions supports the statement of Pielke et al. (2007).

There is, if at all, a local balance of energy flux densities in which the so-called net radiation (also called the radiation balance),

$$R_B = R_{S\downarrow} \left(1 - \alpha_S\right) + \varepsilon\, R_{L\downarrow} - \varepsilon\, \sigma\, T_s^4 \quad , \qquad (1.4)$$

is included. Here, only bare soil is considered for the purpose of simplification, where $R_{S\downarrow}$ is the global (direct plus diffusive solar) radiation, $\alpha_S$ the albedo of the short-wave range, $R_{L\downarrow}$ is the incoming long-wave radiation emitted by constituents of the atmosphere in finite spectral ranges, $\varepsilon = 1 - \alpha_L$ is the absorptivity that is equal to the emissivity, $\alpha_L$ is the albedo of the long-wave range, $\sigma$ is the Stefan constant, and $T_s$ is the surface temperature. Note that the use of the power law of Stefan (1879) and Boltzmann (1884) requires a local formulation because its derivation is not only based on the integration of Planck's (1901) blackbody radiation law, for instance, over all frequencies (from zero to infinity), but also on the integration of the isotropic emission of radiant energy by a small spot (like a hole in the opaque walls of a cavity) over the adjacent half space (e.g., Liou, 2002; Kramm and Mölders, 2009). We may assume that the condition of the local thermodynamic equilibrium is fulfilled (up to 60 km or so above ground). Furthermore, a flux density (hereafter simply denoted as a flux) is counted positive when it is directed to the Earth's surface.

To determine the temperature and the humidity at the Earth's surface in the case of bare soil[1] the balance equations at the Earth's surface of energy fluxes (e.g., Lettau, 1969; Kramm 2007),

$$F_1(\mathbf{S}) = R_B - L_v(T_s)\,Q - H + G = 0 \quad , \tag{1.5}$$

and water fluxes,

$$F_2(\mathbf{S}) = P - R_O - Q - I = 0 \quad , \tag{1.6}$$

are considered. Here, $L_v(T_s)$ is the specific heat of specific heat of phase transition (e.g., vaporization, sublimation), considered as dependent on $T_s$. Moreover, Q and H are the fluxes of water vapor and sensible heat within the atmosphere caused by mainly molecular effects in the immediate vicinity of the Earth's surface and by turbulent effects in the layers above, G is the soil heat flux, P is the precipitation, $R_O$ is the surface runoff, and I is the infiltration. This coupled set of simple equations already documents the difficulty and challenge related to the prediction of second kind. The net radiation not only depends on the parameterization of the fluxes of sensible heat and water vapor and the soil heat flux, but also on the parameterization of surface runoff and infiltration and the prediction of precipitation as a result of the simulated cloud microphysical processes. These processes also affect the radiation transfer of both solar and terrestrial radiation as well as the surface properties like the integral values of the shortwave albedo and the emissivity. Even though parameterization schemes are indispensable in numerical modeling of the atmosphere we have to recognize that any kind of parameterization includes a notable degree of uncertainty.

For determining the temperature and the specific humidity at the surface the set of coupled flux balance equations for the energy and water fluxes (1.5) and (1.6) has to be simultaneously solved with respect to the temperature, $T_s$, and the volumetric water content, $W_s$, at the surface, where customarily a Newton-Raphson iteration procedure of first order,

---

[1]) The inclusion of a vegetation canopy has been discussed, for instance, by Deardorff (1978), McCumber (1980, see also Pielke, 1984), Meyers and Paw U (1986, 1987), Sellers et al. (1986), Braud et al. (1995), Kramm et al. (1996a), Kramm et al. (1998), Ziemann (1998), Su et al. (1998, 2000), Pyles et al. (2000, 2003), and Mölders et al. (2003a,b).

$$\mathbf{S}^{(n+1)} = \mathbf{S}^{(n)} - DF\left(\mathbf{S}^{(n)}\right)^{-1} F\left(\mathbf{S}^{(n)}\right) \quad , \tag{1.7}$$

is used (e.g., Pielke, 1984; Kramm et al., 1996a; Mölders, 1999; Mölders et al., 2003). Here, $\mathbf{S} = \left\{T_s, W_s\right\}^T$ and $F\left(\mathbf{S}\right) = \left\{F_1\left(\mathbf{S}\right), F_2\left(\mathbf{S}\right)\right\}^T$. Furthermore, $DF\left(\mathbf{S}\right)$ is the functional matrix, $DF\left(\mathbf{S}\right)^{-1}$ is the corresponding inverse matrix, and the superscript $T$ denotes the transpose. For this purpose the fluxes of sensible heat and water vapor (see Eqs. (2.15) and (2.16)), and the heat and water fluxes within the soil have to be related to the surface values of temperature and specific humidity. The latter may be expressed as a function of the volumetric water content, $W$.

The coupled set of balance equations (1.5) and (1.6) can also be applied to derive Lettau's (1969) climatonomy equation. Lettau (1969) argued that for long-term considerations, the soil heat flux, G, and the infiltration, I, are of minor importance, and, hence, negligible. Following this argument, one obtains (Kramm, 2007)

$$Bu = \frac{R_B}{L_v\left(T_s\right) P} = \left(1 - A_R\right)\left(1 + Bo\right) \tag{1.8}$$

that may serve as a physically based measure for assessing local and regional climates. Here, Bu is the dryness index (Budyko, 1958) also called the Budyko ratio. It relates the radiation balance to the portion of energy that is necessary to vaporize precipitation completely. Since the term $L_v\left(T_s\right) P$ is always positive, the dryness index has the same sign as the radiation balance. The other quantities are the Bowen ratio, $Bo = H\big/\left(L_v\left(T_s\right) Q\right)$, and the runoff ratio, $A_R = R_O/P$. As Flohn (1988) pointed out, Lettau's climatonomy equation may serve as a suitable tool to characterize arid regions (i.e., drought). Charts of Bu-values, for instance, were plotted by Henning and Flohn and served as their contribution to the desertification conference in 1974 (as cited by Flohn, 1988).

In comparison with bare soil, the determination of the temperature at water surfaces is more complex because (a) a fraction of the incident solar radiation may penetrate the water to a considerable depth without significant absorption, and (b) at both sides of the atmosphere-water interface the transition of a viscous transfer to a fully turbulent transfer has to be considered (see Figure 3).

Since the prediction of climate is mainly considered as a prediction of second kind, it is indispensable to assess the accuracy with which these boundary conditions can be determined so that we can find a reasonable answer, whether climate is predictable with a sufficient degree of accuracy or not.

In our contribution we will discuss the accuracy with which the fluxes of sensible heat and water vapour can be parameterized. These fluxes are required for solving the equation set (1.5) and (1.6) for predicting the lower boundary conditions in numerical models of the atmosphere. The results presented here are mainly based on the review paper of Kramm and Herbert (2009) focused on the physics of the atmospheric surface layer (ASL). In addition, flux aggregation principles indispensable for considering landscapes of patchy fields are debated.

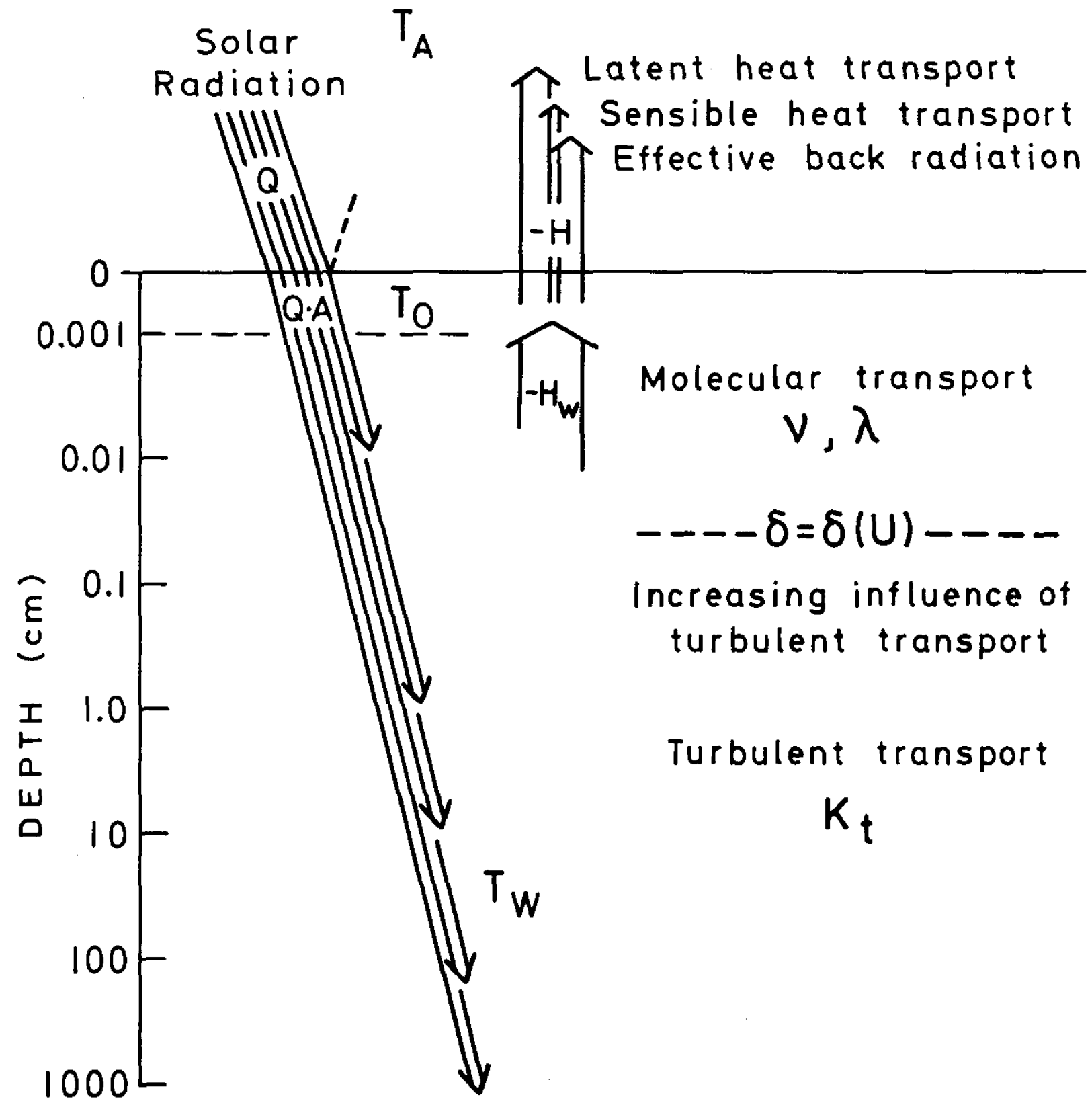

**Figure 3:** Schematic diagram of the heat flow at the air-sea interface (adopted from Hasse, 1971). Note that $Q$ is the incident solar radiation an a horizontal surface, $A$ is the fraction of this radiation penetrating into the water, $\delta = \delta(U)$ (depending on the horizontal wind speed $U$) is the depth of the water layer mainly governed by molecular effect, $T_0$ is the representative temperature of the water skin, $T_W$ is the water temperature, and $H_W$ is the heat flux within the water. Furthermore, $K_t$ is the eddy diffusivity, and $\nu$ and $\lambda$ are the kinematic viscosity and the molecular diffusivity of water, respectively.

## 2. On the parameterization of the fluxes of momentum, sensible heat and water vapor.

Customarily, the fluxes of sensible heat and water vapor are parameterized by addressing the transfer processes in the interfacial sublayer in the immediate vicinity of the Earth's surface and the fully turbulent layer above.

### *2.1. The transfer in the interfacial sublayer*

If we assume horizontally homogeneous and steady-state conditions, we may express the respective flux components in the interfacial sublayer by

Momentum:

$$\tau = \left| \boldsymbol{\tau} \right| = \bar{\rho}\, u_* \left( \frac{\xi_d}{2} \right)^{\frac{1}{2}} \left( \widehat{u}_r - \widehat{u}_s \right) = \bar{\rho}\, u_*^{\,2} = \text{const.} \tag{2.1}$$

Sensible heat:

$$H = -\, c_{p,d}\, \bar{\rho}\, u_*\, B_h \left( \widehat{T}_r - \widehat{T}_s \right) = -\, c_{p,d}\, \bar{\rho}\, u_*\, \Theta_* = \text{const.} \tag{2.2}$$

Water vapor:

$$Q = -\, \bar{\rho}\, u_*\, B_q \left( \widehat{q}_r - \widehat{q}_s \right) = -\, \bar{\rho}\, u_*\, q_* = \text{const.} \tag{2.3}$$

Here, $\rho$ is the air density, u∗ is the friction velocity defined by $u_* = \sqrt{\tau / \bar{\rho}}$, where $\tau$ is the magnitude of the friction stress vector, $\boldsymbol{\tau}$, $\widehat{u}_r$ and $\widehat{u}_s$ are the mean values of the flow velocity at the outer edge of the interfacial sublayer (subscript r), and at the surface (subscript s), where in the case of rigid walls, as considered here, the latter is equal to zero. Furthermore, $\widehat{T}_r$ and $\widehat{T}_s$ are the corresponding mean values of the absolute temperature, and $\widehat{q}_r$ and $\widehat{q}_s$ are those of the specific humidity, respectively. Moreover, $c_{p,d}$ is the specific heat of dry air at constant pressure, $\xi_d = 2\left(u_* / \overline{u}_r\right)^2$ is the local drag coefficient, and $B_{h,q}$ are the sublayer-Stanton number (subscript h) and the sublayer-Dalton number for water vapor. These characteristic numbers may be considered as functions of the local (or roughness) Reynolds number defined by $\eta = u_* \left(z - d\right) / \nu$, the Prandtl number $Pr = \nu / \alpha_T$, and the Schmidt number $Sc_q = \nu / D_q$ for water vapor, respectively, where, $\nu$ is the kinematic viscosity, $z$ is the height above the bare soil surface, $\alpha_T$ is the thermal diffusivity, and $D_q$ is the corresponding molecular diffusivity. Note that Hesselberg's (1926) density-weighted average, $\widehat{\psi} = \overline{\rho\, \psi} / \bar{\rho}$, is denoted by a hat, where $\psi$ represents a field quantity like $u$, $\Theta$, and $q$, while the overbar $(\overline{\ldots})$ designates the Reynolds' (1895) mean; the deviations from Reynolds' mean and Hesselberg's mean are denoted by a prime (') and a double prime ("), respectively. Furthermore, a Hesselberg mean can be related to that of Reynolds by (e.g., van Mieghem, 1973; Cox, 1995; Kramm et al., 1995; Herbert, 1995)

$$\widehat{\psi} = \bar{\psi} + \frac{\overline{\rho'\, \psi'}}{\bar{\rho}} = \bar{\psi} \left\{ 1 + \frac{\overline{\rho'\, \psi'}}{\bar{\rho}\, \bar{\psi}} \right\} \quad , \tag{2.4}$$

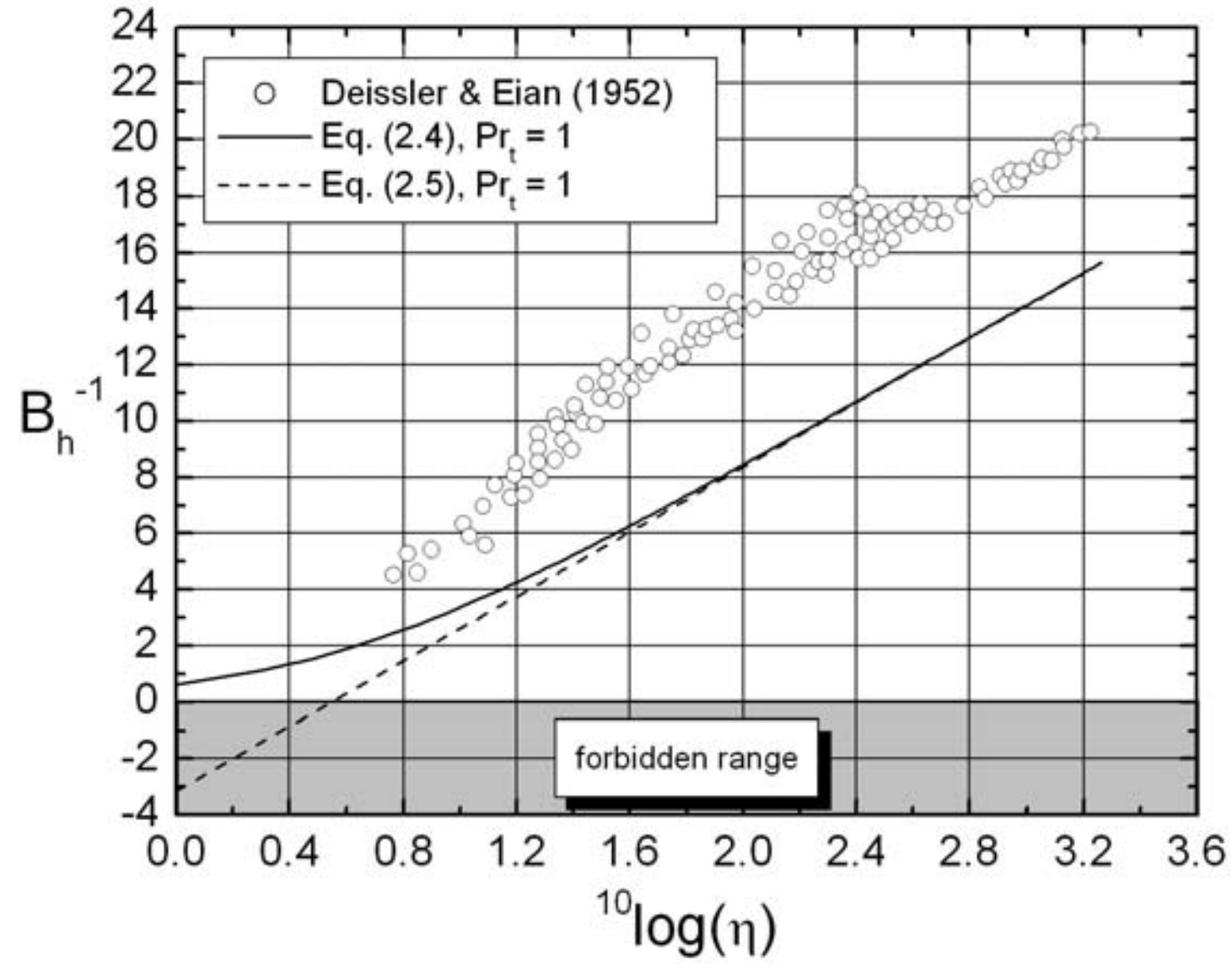

**Figure 4:** Sublayer-Stanton number profiles for the interfacial sublayer over aerodynamically smooth surfaces, where Sheppard's (1958) $K_m/\nu$-approach is considered, and compared with the laboratory data of Deissler and Eian (1952) (adopted from Kramm et al., 2002).

where $\hat{\psi}$ and $\bar{\psi}$ are nearly identical if $\left|\overline{\rho' \psi'}\right| / \left\{\bar{\rho}\, \bar{\psi}\right\} \ll 1$.

The sublayer Stanton number and the sublayer Dalton number are commonly expressed by

$$B_i^{-1} = \frac{1}{\kappa} \ln\left(1 + \frac{z_r}{z_{\chi,i}}\right) \tag{2.5}$$

where $z_{\chi,i} = D_i / (\kappa\, u_*)$ is the so-called roughness length for a scalar quantity like sensible heat (i = h) and water vapor (i = q), $D_i$ stands for the molecular diffusivity of water vapor and the thermal diffusivity, and $\kappa$ is the von Kármán constant. As shown by Kramm et al. (1995), Eq. (2.5) is based on Sheppard's (1958) formulation of an effective diffusivity $K_{eff,i} = D_i + u_*\, \kappa\, z$. This formulation of an effective diffusivity is based on an oversimplification for very small values of $z$, as already assumed by Sheppard (1958). Unfortunately, formula (2.5) is customarily approximated by

$$B_i^{-1} = \frac{1}{\kappa} \ln\left(\frac{z_0}{z_{\chi,i}}\right) \quad , \tag{2.6}$$

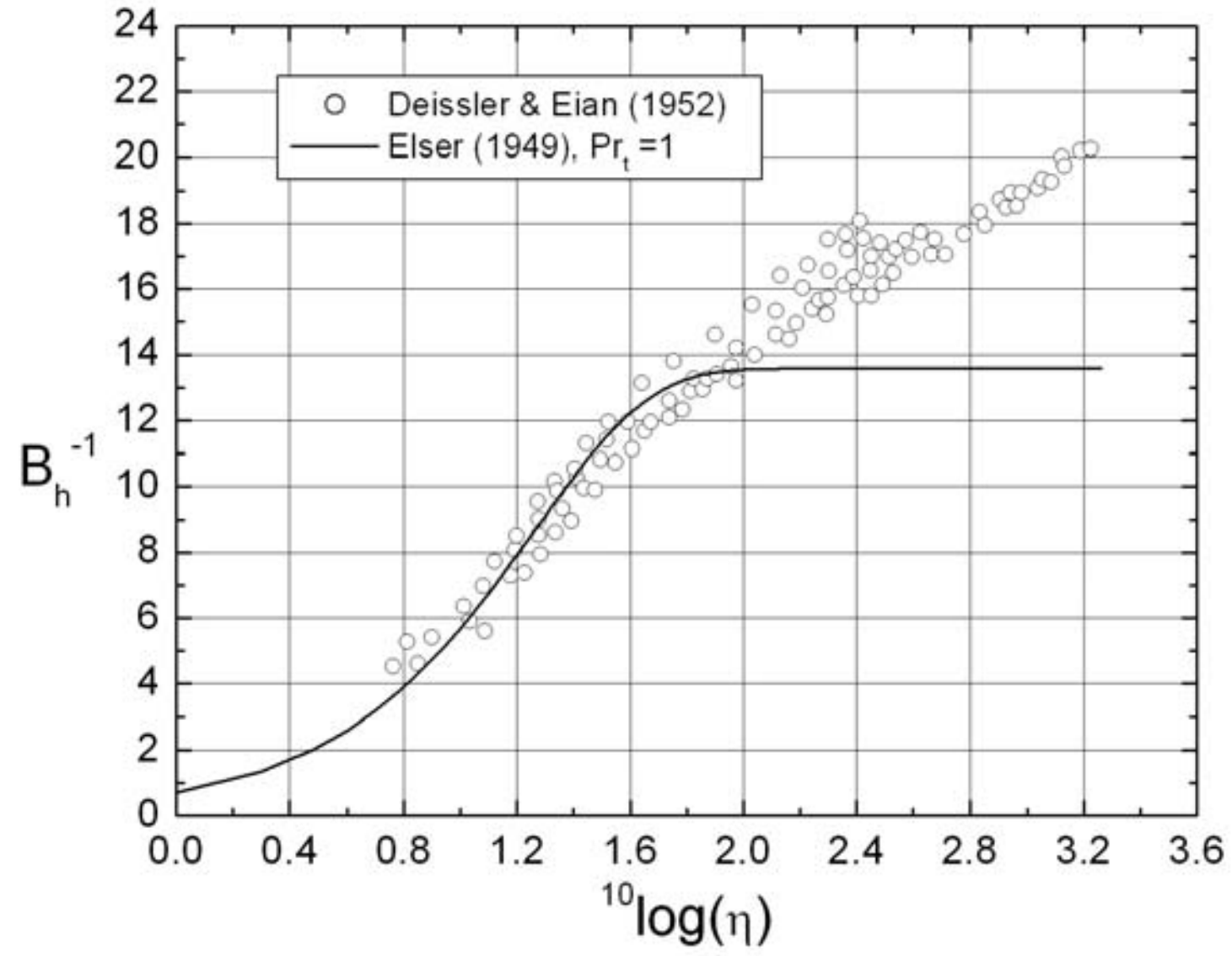

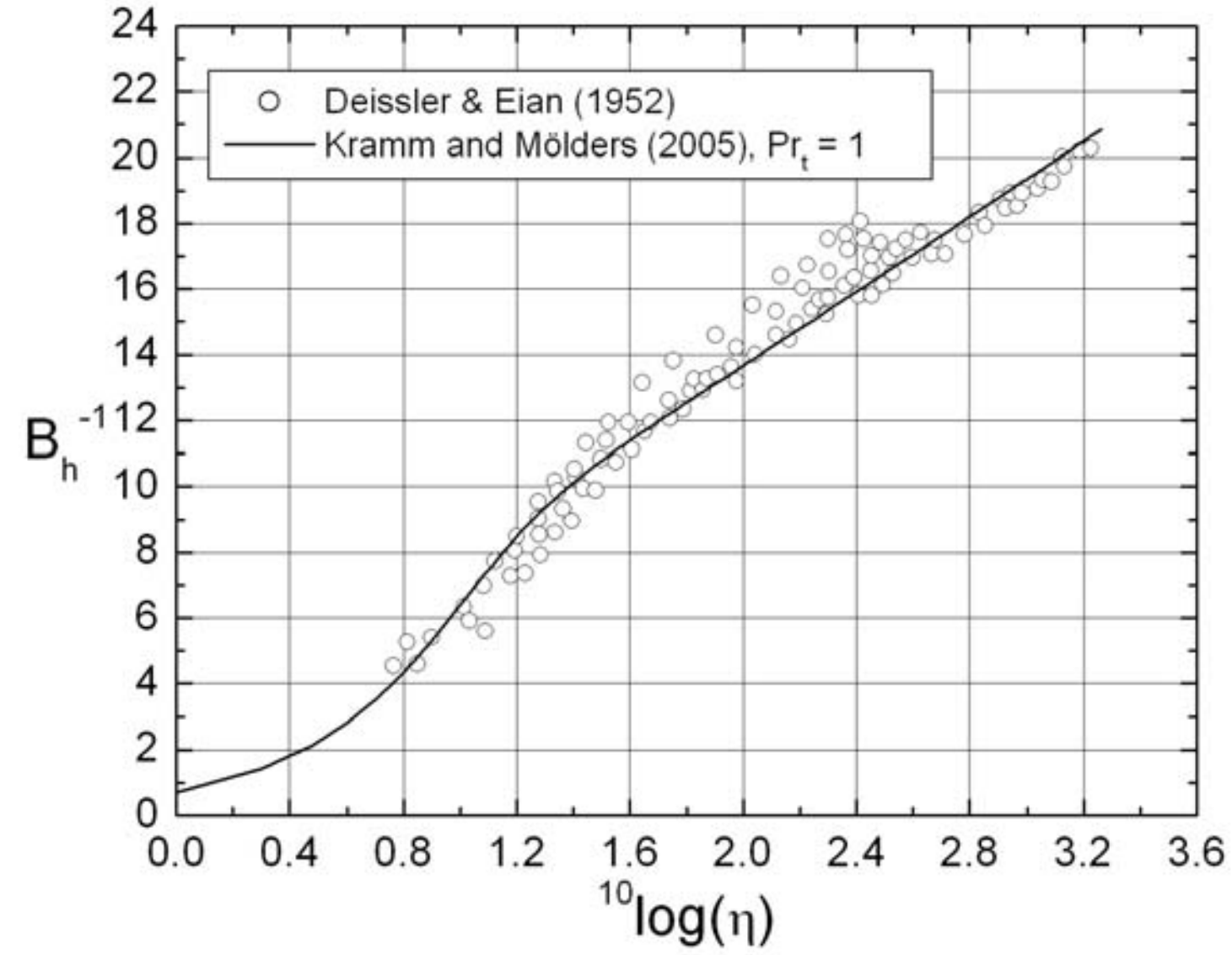

**Figure 5:** As in Figure 4, but for Elser (1949, above) and Kramm and Mölders (2005, below).

i.e., the roughness length $z_{\chi,i}$ is considered as a level at which a scalar quantity should take its surface value if the logarithmic law would be valid down to the surface (e.g., Chamberlain, 1966, 1968; Garratt and Hicks, 1973; Brutsaert, 1975). However, formula (2.6) is only acceptable, if at all, for $z_0 \gg z_{\chi,i}$ (Kramm et al., 1995). Here, $z_0$ is the roughness length for momentum. It defines the height at which the mean wind speed extrapolates to zero when the height above ground is approaching to $z_0$ (or $z_0 + d$, when a zero-plane displacement, $d$, is introduced additionally). Obviously, this definition clearly differs from that of the roughness length of a scalar quantity. Thus, $z_{\chi,i}$ cannot be considered as an analog to $z_0$. As suggested by Garratt and Hicks (1973) for practical purposes, an uppermost value of $\kappa\, B_i^{-1} = \ln\left(z_0 / z_{\chi,i}\right) \approx 2.3$ $\Rightarrow\; z_0 / z_{\chi,i} \approx 10$ is customarily used, especially in attempts to determine fluxes of sensible and latent heat from satellite equipments (e.g., Su et al., 1999). However, as documented by Kramm and Herbert (1984), Beljaars and Holtslag (1991), Müller et al. (1993), and Kramm et al. (1996b) $\kappa\, B_i^{-1}$-values much larger than 2.3 exist for surfaces covered by vegetation. Meanwhile, there are awkward attempts to determine the roughness lengths for scalar quantities in a similar way than that for momentum (e.g., Blyth and Dolman, 1995), although the former one was introduced by Sheppard (1958) and Garratt and Hicks (1973) by the definition $z_{\chi,i} = D_i / \left(\kappa\, u_*\right)$.

If in the case of aerodynamically smooth surfaces like water and ice surfaces $B_i^{-1}$ is approximated by formula (2.6), the positive-definite sublayer Stanton number might become either zero or negative because $z_0 \le z_{\chi,i}$ (see Figure 4). If the transfer of sensible heat and matter over water surfaces is described using the so-called surface renewal model of Liu et al. (1979), such an inconsistent sublayer Stanton number may not occur. This model completely agrees with that of Elser (1949) for the viscous sublayer and the transition layer, i.e., it does not satisfy the equation of continuity (e.g., Reichardt (1950, 1951; Elrod, 1957; Hinze, 1959; Monin and Yaglom, 1971; Kramm et al., 2002; Kramm and Mölders, 2005). Results for the parameterisation schemes of Elser (1949) and Kramm and Mölders (2005) are illustrated in Figure 5 for the purpose of comparison. Obviously, in the case of Elser's one it is indispensable to combine it with an approach for a turbulent flow where the matching point is of about $\eta = 40$ (see also Liu et al, 1979).

### *2.2. The transfer in the fully turbulent ASL*

#### 2.2.1 Some introducing remarks

There are local balance equations (also called conservation equations) for momentum (Newton's 2nd axiom), internal energy (1st law of thermodynamics), total energy, atmospheric constituents, and total mass (equation of continuity). In his famous article, Vilhelm Bjerknes (1904) also listed the balance equation for entropy (2nd law thermodynamics), but since in short-term integration like weather prediction for the next couple of days irreversible processes may not play a dominant role. However, we do not know its influence in long-term integrations like climate projections.

The balance equation mentioned before were derived for macroscopic (molecular) systems, but important layers of the atmosphere (and the oceans) are in a turbulent state which means that stochastic motions have to be described. To solve these balance equations for turbulent systems we usually follow the guidelines of Osborne Reynolds (1895) because we know that these guidelines lead to reasonable results in aerodynamics and fluid mechanics. Introducing statistical elements into the original balance equations leads to a loss of exactness because we now have more unknowns than balance equations (closure problem of turbulence). And these unknowns have to be related to known variables so that the full set of balance equations can be solved. We called it the parameterization. However, as already mentioned, any kind of parameterization includes a notable degree of uncertainty. The balance equation of the total enthalpy (it means the enthalpy plus the turbulent kinetic energy), for instance, may be applied to assess such parameterization schemes for turbulent fluids in long-term predictions (e.g., Kramm and Meixner, 2000).

### 2.2.2 Flux-profile relations for the ASL

Under horizontally homogeneous and steady-state conditions the corresponding flux components in the fully turbulent ASL read (Kramm and Herbert, 2009)

Momentum:

$$\tau = |\boldsymbol{\tau}| = \left( \overline{\rho\, u''\, w''}^{2} + \overline{\rho\, v''\, w''}^{2} \right)^{\frac{1}{2}} = \bar{\rho}\, u_*^{\,2} = \text{const.} \quad , \tag{2.7}$$

Sensible heat:

$$H = c_{p,d}\, \overline{\rho\, w''\, \Theta''} = -\, c_{p,d}\, \bar{\rho}\, u_*\, \Theta_* = \text{const.} \quad , \tag{2.8}$$

Water vapor:

$$Q = \overline{\rho\, w''\, q''} = -\, \bar{\rho}\, u_*\, q_* = \text{const.} \tag{2.9}$$

These covariance terms occur as the result of two steps: (a) in the corresponding macroscopic balance equations the instantaneous field quantities are expressed by a mean value and the deviation from that, (b) these macroscopic balance equations are subsequently averaged (Reynolds, 1896). Here, $\Theta_*$ is the heat flux temperature (also called the temperature scale), $q_*$ is the water vapor flux concentration (also called the humidity scale). Equations (2.7) to (2.9) reflect a prominent prerequisite in micrometeorology, namely that the friction stress vector and the vertical components of the turbulent fluxes of heat and water vapor (here designated as micrometeorological fluxes) are invariant with height. From a mathematical point of view it can be expressed by $\partial F / \partial z = 0 \Leftrightarrow F = \text{const}(z)$, where F stands for the micrometeorological fluxes of momentum (i.e., the magnitude of the friction stress vector), sensible heat and water vapor.

This height-invariance of the micrometeorological fluxes may serve to define the thickness of the ASL (see Kramm and Herbert, 2009). As outlined before, it generally demands that steady-state conditions and the condition of horizontally uniform fields of mean wind speed (i.e., the mean vertical velocity component is equal to zero), mean temperature, and mean humidity are fulfilled. In addition, net source and sink effects owing to phase transition processes are excluded. Even though the condition of height invariance may customarily be fulfilled only in a micrometeorological sense (i.e., these micrometeorological fluxes may vary with height across the entire ASL, but not more than 10 percent of their values in the immediate vicinity of the surface), it serves as the basis for the so-called constant flux approximation on which micrometeorological scaling is based, namely (a) Monin-Obukhov scaling for forced-convective conditions (Monin and Obukhov, 1954), and (b) Prandt-Obukhov-Priestley scaling for free-convective conditions (Prandtl, 1932; Obukhov, 1946; Priestley, 1959), respectively. The latter may be considered as asymptotic ones for force-convective conditions.

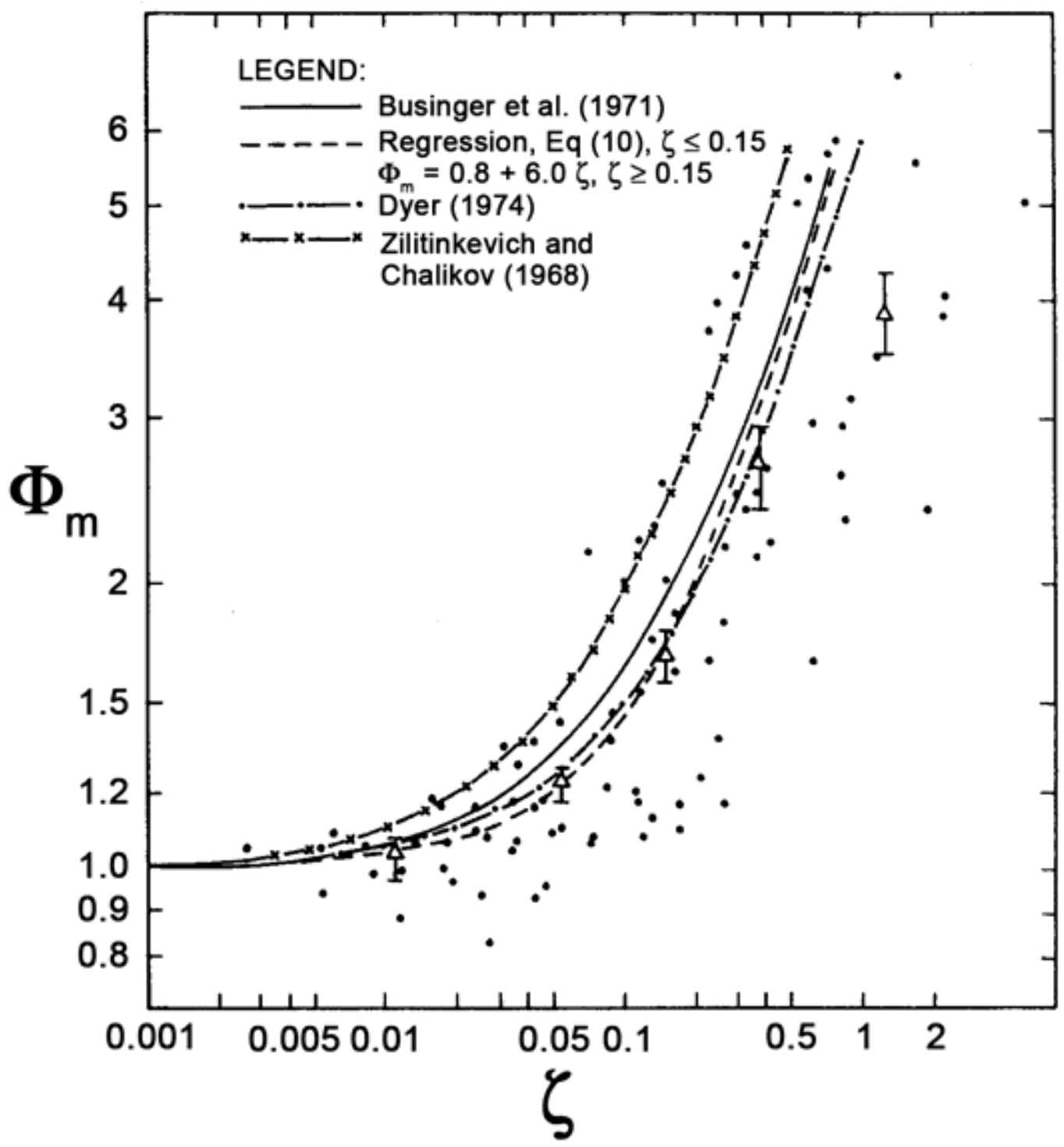

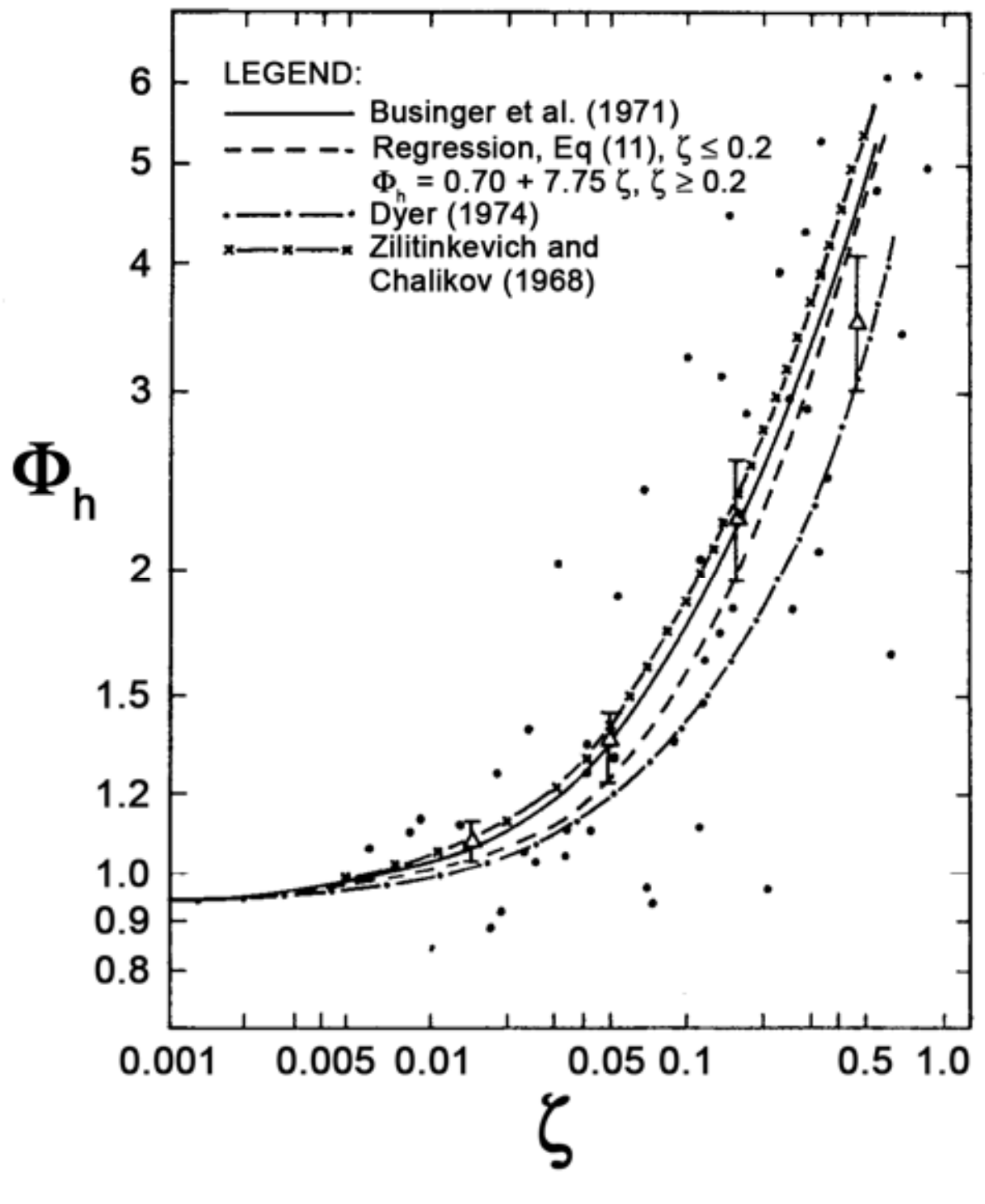

**Figure 6**: The local similarity functions $\Phi_m(\zeta)$ and $\Phi_h(\zeta)$ as a function of $\zeta$ for stable stratification (adopted from Högström, 1988).

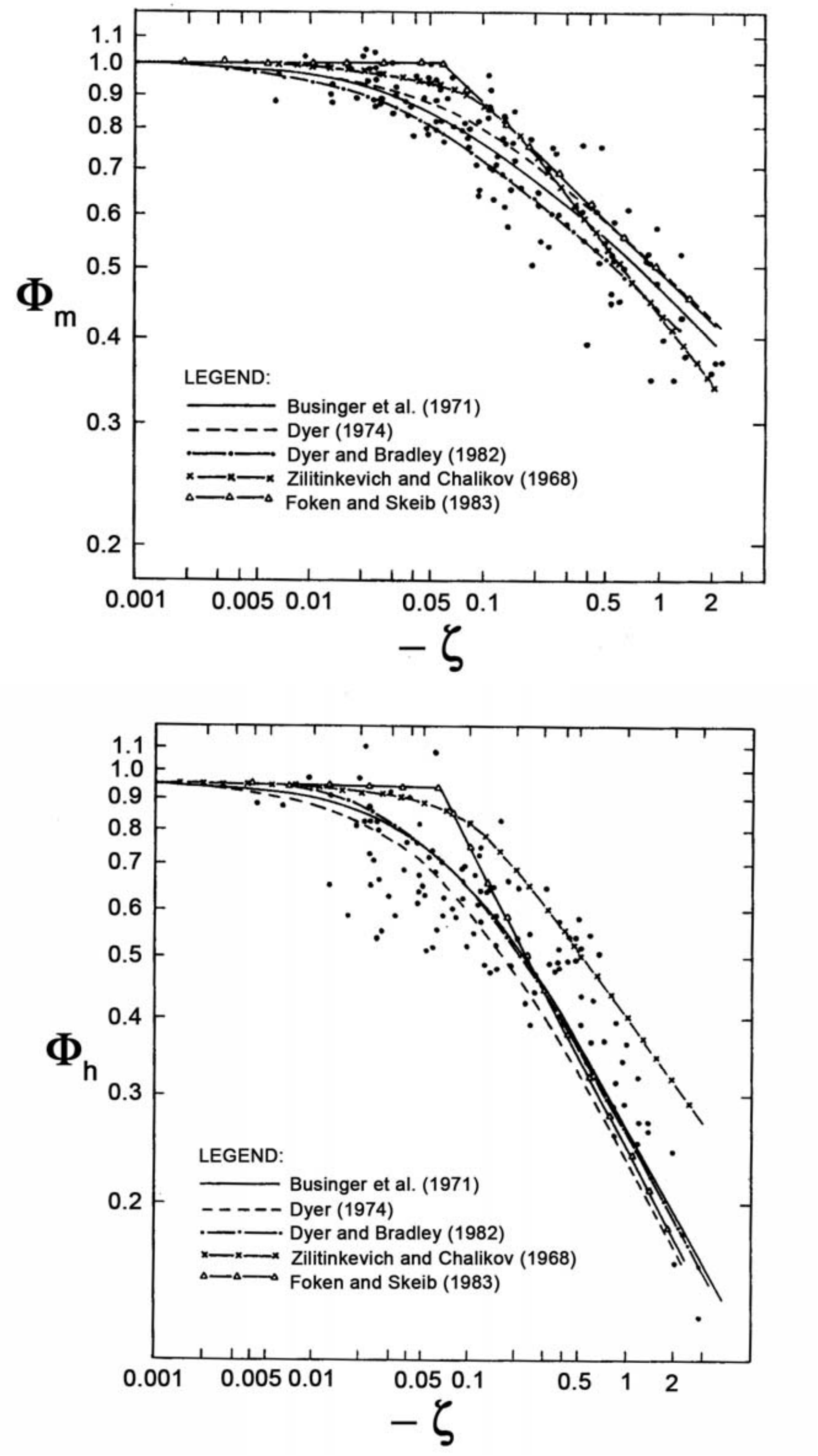

**Figure 7**: As in Figure 6, but for unstable stratification.

The friction velocity along with these scaling quantities can be related to measured or predicted mean values of the vertical profiles of mean wind, potential temperature, and specific humidity by (e.g., Panofsky, 1963; Kramm and Herbert, 2009)

$$\widehat{u_R} - \widehat{u_r} = \frac{u_*}{\kappa}\left(\ln\frac{z_R - d}{z_r - d} - \Psi_m\left(\zeta_R, \zeta_r\right)\right) \quad , \tag{2.10}$$

$$\widehat{\Theta_R} - \widehat{\Theta_r} = \frac{\Theta_*}{\kappa}\left(\ln\frac{z_R - d}{z_r - d} - \Psi_h\left(\zeta_R, \zeta_r\right)\right) \quad , \tag{2.11}$$

and

$$\widehat{q_R} - \widehat{q_r} = \frac{q_*}{\kappa}\left(\ln\frac{z_R - d}{z_r - d} - \Psi_q\left(\zeta_R, \zeta_r\right)\right) \quad , \tag{2.12}$$

where the so-called integral similarity (or stability) functions are defined by (Panofsky, 1963)

$$\Psi_{m,h,g}\left(\zeta_R, \zeta_r\right) = \int_{z_r}^{z_R} \frac{1 - \Phi_{m,h,q}\left(\left(z - d\right)/L\right)}{z - d}\,dz = \int_{\zeta_r}^{\zeta_R} \frac{1 - \Phi_{m,h,q}\left(\zeta\right)}{\zeta}\,d\zeta \quad . \tag{2.13}$$

Equations (2.1) to (2.3) and (2.10) to (2.13) can be combined to predict the fluxes of momentum, sensible heat, and water vapor on the basis of the mean quantities of wind speed, temperature and specific humidity at the surface and the reference height $z_R$ (e.g., the height of the ASL). One obtains (e.g., Pal Arya, 1988):

$$\tau = \bar{\rho}\, C_m \left(\widehat{u_R} - \widehat{u_s}\right)^2 = \text{const.} \quad , \tag{2.14}$$

$$H = -\,\bar{\rho}\, c_{p,d}\, C_h \left(\widehat{u_R} - \widehat{u_s}\right)\left(\widehat{\Theta_R} - \widehat{T_s}\right) = \text{const.} \quad , \tag{2.15}$$

and

$$Q = -\,\bar{\rho}\, C_q \left(\widehat{u_R} - \widehat{u_s}\right)\left(\widehat{q_R} - \widehat{q_s}\right) = \text{const.} \quad , \tag{2.16}$$

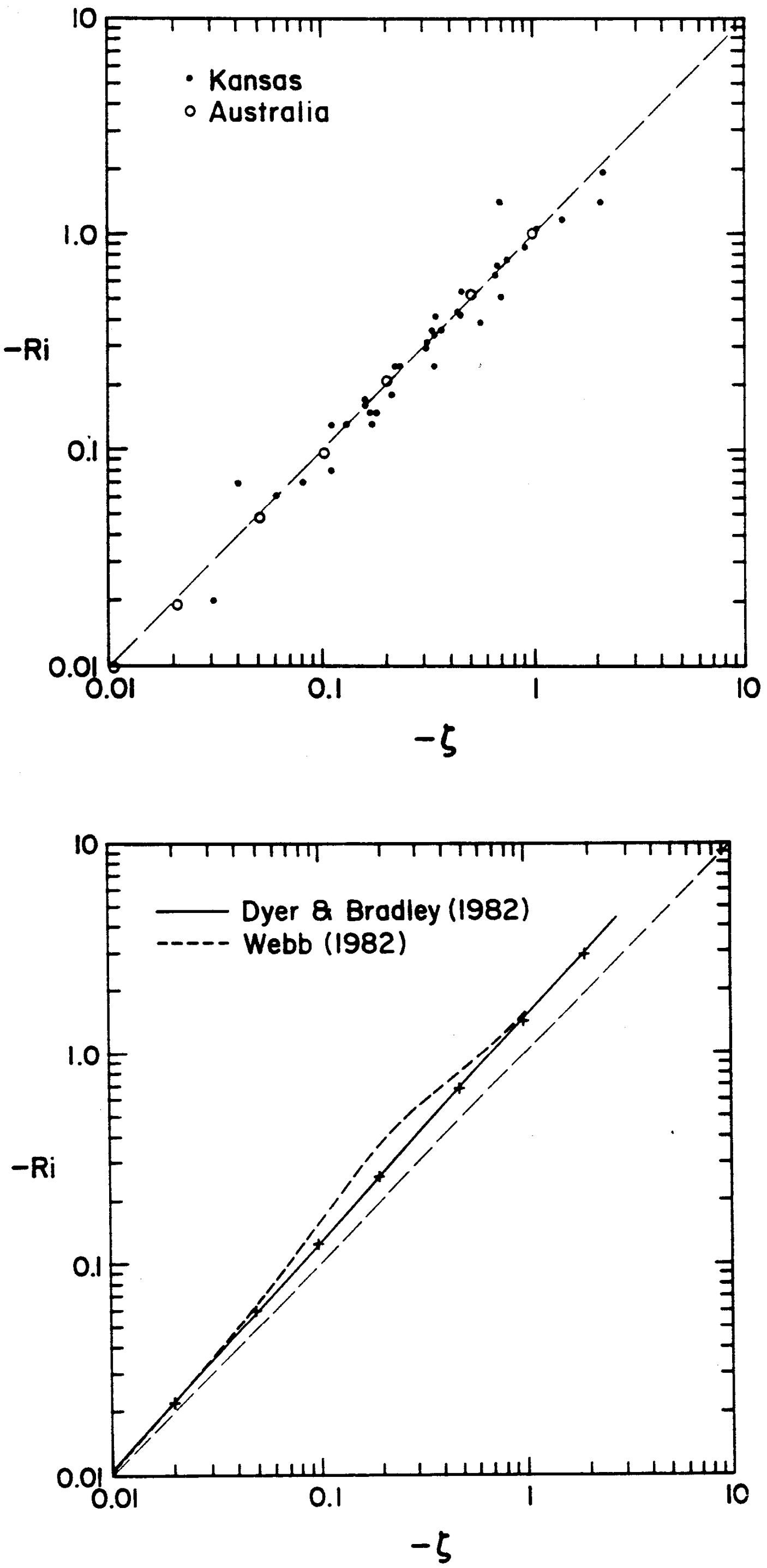

**Figure 8**: Gradient Richardson number, Ri, versus Obukhov number, $\zeta$, for various field experiments (adopted from Businger, 1988).

where $C_m$ is the drag coefficient and $C_h$ and $C_q$ are the transfer coefficients for sensible heat and water vapor given by

$$C_m = \frac{\kappa^2}{\left( \kappa \left( \frac{\xi_d}{2} \right)^{-\frac{1}{2}} + \ln \frac{z_R - d}{z_r - d} - \Psi_m\left(\zeta_R, \zeta_r\right) \right)^2} \quad , \tag{2.17}$$

$$C_h = \frac{\kappa^2}{\left( \kappa \left( \frac{\xi_d}{2} \right)^{-\frac{1}{2}} + \ln \frac{z_R - d}{z_r - d} - \Psi_m\left(\zeta_R, \zeta_r\right) \right) \left( \kappa \, B_h^{-1} + \ln \frac{z_R - d}{z_r - d} - \Psi_h\left(\zeta_R, \zeta_r\right) \right)} \quad , \tag{2.18}$$

and

$$C_q = \frac{\kappa^2}{\left( \kappa \left( \frac{\xi_d}{2} \right)^{-\frac{1}{2}} + \ln \frac{z_R - d}{z_r - d} - \Psi_m\left(\zeta_R, \zeta_r\right) \right) \left( \kappa \, B_q^{-1} + \ln \frac{z_R - d}{z_r - d} - \Psi_q\left(\zeta_R, \zeta_r\right) \right)} \quad . \tag{2.19}$$

Note that in the case of Eq. (2.15) the assumption $\widehat{\Theta_r} = \widehat{T_r}$ was used. Furthermore, in the case of rigid surfaces the velocity $\widehat{u_s}$ is equal to zero. Moreover, long-lived trace constituents like carbon dioxide ($CO_2$) may be handled like water vapor and sensible heat.

The quantities $\Phi_m\left(\zeta\right)$, $\Phi_h\left(\zeta\right)$, and $\Phi_q\left(\zeta\right)$ are the local similarity functions for momentum, sensible heat, and water vapor established by the similarity hypotheses of Monin and Obukhov (1954)[2],

$$\frac{\kappa\left(z - d\right)}{u_*} \frac{\partial \hat{u}}{\partial z} = \Phi_m\left(\zeta\right) \quad , \tag{2.20}$$

$$\frac{\kappa\left(z - d\right)}{\Theta_*} \frac{\partial \widehat{\Theta}}{\partial z} = \Phi_h\left(\zeta\right) \quad , \tag{2.21}$$

[2]) Often it is called the Monin-Obukhov similarity theory, but this is a misnomer. The theory behind the similarity hypotheses of Monin and Obukhov is dimensional analysis as introduced into the literature by Buckingham (1914).

and

$$\frac{\kappa\left(z-d\right)}{q_*}\frac{\partial\hat{q}}{\partial z}=\Phi_q\left(\zeta\right) \tag{2.22}$$

which can also be considered as non-dimensional gradients. From the perspective of dimensional analysis these $\Phi$-functions may be denoted as universal functions. Here, $\zeta=\left(z-d\right)/L$ is the Obukhov number, and $L$ is the Obukhov stability length given by (Obukhov, 1946; Monin and Obukhov, 1954, Zilitinkevič, 1966)

$$L=-\frac{c_{p,d}\,\bar{\rho}\,u_*^{\,3}}{\kappa\,\frac{g}{\hat{\Theta}}\left(H+0.61\,c_{p,d}\,\hat{\Theta}\,Q\right)}=\frac{u_*^{\,2}}{\kappa\,\frac{g}{\hat{\Theta}}\left(\Theta_*+0.61\,\hat{\Theta}\,q_*\right)}\quad. \tag{2.23}$$

Note that the mean lateral wind component, $\hat{v}$, is arbitrarily chosen as equal to zero. This choice can be justified by arranging the x-axis (and, hence, the y-axis) in such a manner that $\hat{v}$ vanishes.

### 2.2.3 Empirical $\Phi$-functions for the transfer of momentum, sensible heat, and water vapor

Since the similarity hypotheses of Monin and Obukhov (1954) can only serve to show that universal functions may exist, such conventional $\Phi$-functions for the transfer of momentum, sensible heat, and water vapor have to be determined empirically and/or theoretically. Unfortunately, the results of these $\Phi$-functions obtained from sophisticated field campaigns show a considerable scatter (see Figures 6 and 7).

The empirical results of Zilitinkevič and Čalikov (1968) for stable stratification and Dyer and Hicks (1970) for unstable stratification, for instance, may be gathered by (e.g., Kramm and Herbert, 2009)

$$\Phi_m\left(\zeta\right)=\begin{cases}\left(1-\gamma_2\,\zeta\right)^{-1/4} & \text{for } \zeta<0 \quad \left(\text{unstable}\right)\\ 1 & \text{for } \zeta=0 \quad \left(\text{neutral}\right)\\ 1+\gamma_1\,\zeta & \text{for } \zeta>0 \quad \left(\text{stable}\right)\end{cases} \tag{2.24}$$

and

$$\Phi_q\left(\zeta\right)=\Phi_h\left(\zeta\right)=\begin{cases}\Phi_m^{\,2}\left(\zeta\right) & \text{for } \zeta<0\\ 1 & \text{for } \zeta=0\\ \Phi_m\left(\zeta\right) & \text{for } \zeta>0\end{cases} \tag{2.25}$$

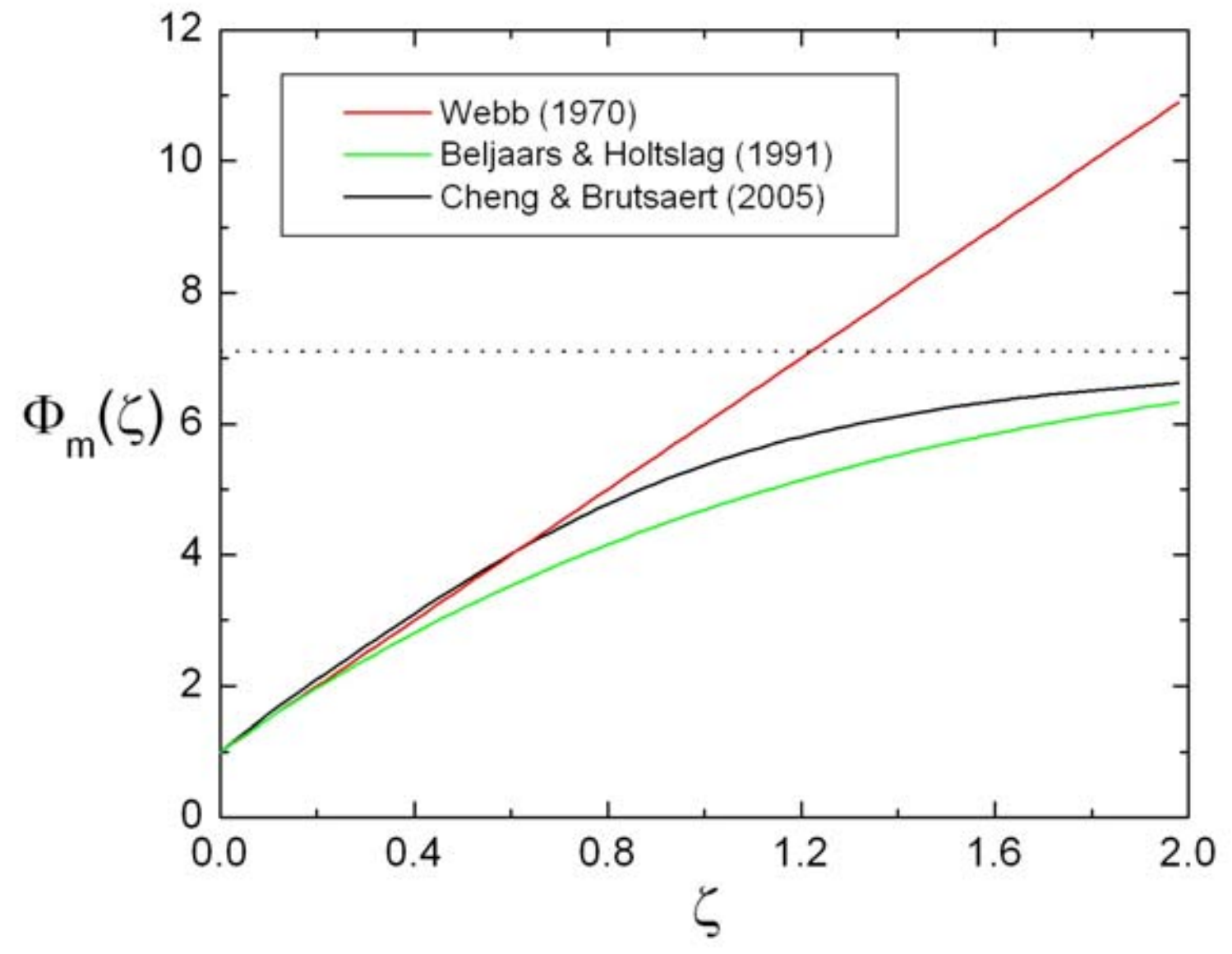

**Figure 9:** The non-dimensional wind shear, $\Phi_m(\zeta)$, as a function of the Obukhov number, $\zeta$, for stable stratification. Shown are the formulae of Webb (1970), Beljaars and Holtslag (1991), and Cheng and Brutsaert (2005). The dotted line characterizes the asymptotic solution for the $\Phi_m(\zeta)$-function of Cheng and Brutsaert (2005) (adopted from Mölders and Kramm, 2009).

with $\gamma_1 \cong 5$ and $\gamma_2 \cong 16$. The relationship $\Phi_m(\zeta) = (1 - \gamma_2\,\zeta)^{-1/4}$ in Eq. (2.24) is called the Businger-Dyer-Pandolfo relationship (Dyer, unpublished; Businger, 1966, 1988; Pandolfo, 1966), later experimentally proved by Dyer and Hicks (1970), Businger et al. (1971) and others, where their results mainly cover the stability range $-2 \le \zeta < 0$ (see also Panofsky and Dutton, 1984; Högström, 1988; Sorbjan, 1989; and Figure 7). The linear formula $\Phi_m(\zeta) = 1 + \gamma_1\,\zeta$ in Eq. (2.24) was first recommended by Monin and Obukhov (1954) for stable stratification (and weakly unstable stratification) and later experimentally proved by Čalikov (1968), Zilitinkevič and Čalikov (1968), Businger et al. (1971) and others mainly for the stability range $0 \le \zeta < 1$, but there is a large scatter in the case of momentum with some values of $\Phi_m(\zeta)$ for $\zeta > 1$ (see Figure 6). The relationship for unstable stratification, $\Phi_h(\zeta) = \Phi_m^{\,2}(\zeta)$ in Eq. (2.25), as already suggested by Businger (1966) and Pandolfo (1966) was eventually proved by Dyer and Hicks (1970) for the stability range $-1 \le \zeta < 0$. As recommended by Webb (1970), the relationship $\Phi_q(\zeta) = \Phi_h(\zeta) = \Phi_m(\zeta)$ may be acceptable for stable stratification.

Expressing the gradient-Richardson number by the non-dimensional gradients and assuming that $\Phi_q(\zeta) = \Phi_h(\zeta)$ (see formula (2.25)) yield then (e.g., Pai Mazumder, 2006)

$$\mathrm{Ri} = \frac{\Phi_h(\zeta)}{\left(\Phi_m(\zeta)\right)^2}\,\zeta \quad . \tag{2.26}$$

Obviously, the Businger-Pandolfo relationship $\Phi_h(\zeta) = \Phi_m{}^2(\zeta)$ for unstable stratification leads to

$$\mathrm{Ri} = \zeta \quad \text{for } \zeta < 0 \quad . \tag{2.27}$$

Dyer and Bradley (1982) and Webb (1982), however, pointed out that small deviations from this identity might occur (see Figure 8).

If we accept Webb's (1970) recommendation $\Phi_q(\zeta) = \Phi_h(\zeta) = \Phi_m(\zeta)$ for stable stratification we will obtain

$$\mathrm{Ri} = \frac{\zeta}{1 + \gamma_1\,\zeta} \quad \text{for } \zeta > 0 \quad . \tag{2.28}$$

This recommendation means that the turbulent Prandtl number, $\mathrm{Pr}_t$, is equal (or close) to unity for stable stratification, and the gradient-Richardson number and the flux-Richardson number are (nearly) identical.

Beside the formulae (2.24) and (2.25), Businger et al. (1971) found

$$\Phi_m(\zeta) = \begin{cases} \left(1 - \gamma_4\,\zeta\right)^{-1/4} & \text{for } \zeta < 0 \\ 1 & \text{for } \zeta = 0 \\ 1 + \gamma_3\,\zeta & \text{for } \zeta > 0 \end{cases} \tag{2.29}$$

and

$$\Phi_h(\zeta) = \begin{cases} 0.74\left(1 - \gamma_5\,\zeta\right)^{-1/2} & \text{for } \zeta < 0 \\ 0.74 & \text{for } \zeta = 0 \quad , \\ 0.74 + \gamma_3\,\zeta & \text{for } \zeta > 0 \end{cases} \tag{2.30}$$

with $\gamma_3 \cong 4.7$, $\gamma_4 \cong 15$, and $\gamma_5 \cong 9$. These authors, however, used a von Kármán constant of $\kappa = 0.35$. Introducing these local similarity functions into formula (2.26) provides

$$\mathrm{Ri} = \begin{cases} 0.74 \left( \dfrac{1 - \gamma_4\, \zeta}{1 - \gamma_5\, \zeta} \right)^{\frac{1}{2}} \zeta & \text{for } \zeta < 0 \\ \dfrac{\zeta}{1 + \gamma_3\, \zeta} \left( 1 - \dfrac{0.26}{1 + \gamma_3\, \zeta} \right) & \text{for } \zeta > 0 \end{cases} . \tag{2.31}$$

Obviously, for unstable stratification we have $|\mathrm{Ri}| < |\zeta|$. In the case of stable stratification the influence of the term $0.26/(1 + \gamma_3\, \zeta)$ becomes weaker and weaker when the Obukhov number increases, i.e., the results inferred from formulae (2.28) and (2.31) only differ slightly for strongly stable conditions.

Two prominent difficulties can be attributed to the use of these parameterization principles: Since $\gamma_1 \cong 5$, the gradient Richardson number has to satisfy the condition $\mathrm{Ri} < 0.2$ because $\zeta = \mathrm{Ri} / (1 - \gamma_1\, \mathrm{Ri})$ (inferred from Eq. (2.28)) would become indeterminate for $\mathrm{Ri} = 0.2$ and negative for $\mathrm{Ri} > 0.2$. The latter is in contradiction to the definition of stable stratification (Mölders and Kramm, 2009).

According to Högström (1988; see Figure 6) and Cheng and Brutsaert (2005), reliable values of the Obukhov number satisfy the condition $0 < \zeta \leq 2$ if stable stratification prevails. This means that in the case of $\zeta = 2$ the gradient Richardson number amounts to $\mathrm{Ri} = 0.182$. As $\mathrm{Ri}$ and the flux Richardson number, $\mathrm{Ri}_f$, are related to each other by $\mathrm{Ri} = \mathrm{Pr}_t\, \mathrm{Ri}_f$ and the turbulent Prandtl number is given by $\mathrm{Pr}_t = \Phi_h(\zeta)/\Phi_m(\zeta)$ Webb's (1970) recommendation $\Phi_q(\zeta) = \Phi_h(\zeta) = \Phi_m(\zeta)$ for stable stratification would lead to $\mathrm{Ri} = \mathrm{Ri}_f$ (e.g., Stull, 1988; Kramm, 1995). This means that the critical flux Richardson number would amount to $\mathrm{Ri}_{f,cr} = 0.182$. This value is much smaller than the theoretical value of $\mathrm{Ri}_{f,cr} = 1$ that characterizes the fact that mechanical gain of turbulent kinetic energy (TKE) equals the thermal loss of TKE so that the turbulent flow becomes increasingly viscous (laminar) due to the dissipation of energy (Kramm and Meixner, 2000). Note that the restriction of $\mathrm{Ri}$ was Louis' (1979) reason to introduce a parametric model with which he artificially enhanced the transfer coefficient for sensible heat for strongly stable stratification to prevent "that once the bulk Richardson number (derived from $\mathrm{Ri}$ using finite differences) exceeds its critical value, the ground becomes energetically disconnected from the atmosphere and starts cooling by radiation at a faster rate than is actually observed".

To prevent such an energetic disconnection Beljaars and Holtslag (1991) introduced

$$\left. \begin{aligned} &\Phi_m(\zeta) = 1 + \zeta \left\{ 0.7 \; + 0.5 \exp(-\,0.35\, \zeta) \left( 1 - 0.35\, \zeta + 5 \right) \right\} \\ &\Phi_h(\zeta) = \Phi_m(\zeta) \end{aligned} \right\} \tag{2.32}$$

As illustrated in Figure 9, it differs from that recommended by Webb (1970). Recently, Cheng and Brutsaert (2005) suggested for stable stratification

$$\Phi_m(\zeta) = 1 + \gamma_6 \left( \frac{\zeta + \zeta^{\gamma_7} \left(1 + \zeta^{\gamma_7}\right)^{\frac{1-\gamma_7}{\gamma_7}}}{\zeta + \left(1 + \zeta^{\gamma_7}\right)^{\frac{1}{\gamma_7}}} \right) \tag{2.33}$$

and

$$\Phi_h(\zeta) = 1 + \gamma_8 \left( \frac{\zeta + \zeta^{\gamma_9} \left(1 + \zeta^{\gamma_9}\right)^{\frac{1-\gamma_9}{\gamma_9}}}{\zeta + \left(1 + \zeta^{\gamma_9}\right)^{\frac{1}{\gamma_9}}} \right) \tag{2.34}$$

with $\gamma_6 = 6.1$, $\gamma_7 = 2.5$, $\gamma_8 = 5.3$, and $\gamma_9 = 1.1$.These formulae should cover the entire range of stable stratification. For neutral conditions, i.e., $\zeta = 0$, one obtains $\Phi_h(0) = \Phi_m(0) = 1$. For moderate stable stratification both formulae can be approximated by linear expressions, $\Phi_m(\zeta) \cong 1 + \gamma_6\, \zeta$ and $\Phi_h(\zeta) \cong 1 + \gamma_8\, \zeta$. For increasing stability formulae (2.33) and (2.34) tend to $\Phi_m(\zeta) = 1 + \gamma_6$ and $\Phi_h(\zeta) = 1 + \gamma_8$ (see Figure 9). Obviously, for the entire range of stable stratification $\Phi_m(\zeta)$ and $\Phi_h(\zeta)$ slightly differ from each other. In contrast to the $\Phi_m(\zeta)$-function of Beljaars and Holtslag (1991) for stable stratification, which has a point of inflection at $\zeta \cong 3$, the formulae of Cheng and Brutsaert (2005) are bounded.

The results for strongly stable stratification should generally be considered with care. As reported by Cheng and Brutsaert (2005), the calculated $\Phi_h(\zeta) - 1$ data points for $\zeta > 2$ were excluded from the analysis because the larger scatter suggested either unacceptable error in the measurements or perhaps other unexplained physical effects. As these authors pointed out, one possible reason could be that these data points are already outside the stable surface layer so that Monin-Obukhov similarity, as expressed, for instance, by Eq. (2.21) may not be valid.

It is obvious that formulae (2.33) and (2.34) lead to logarithmic profiles for neutral and strongly stable conditions. The latter, already found by Webb (1970), seems to be awkward because if the magnitude of turbulent fluctuations decreases towards the small values of the quiet regime with increasing stability (e.g., Okamoto and Webb, 1970; Kondo et al., 1978), the near-surface flow should become mainly laminar. In the case of a pure laminar flow viscous effects are dominant leading to $\partial\hat{u}/\partial\eta = u_*$, $\partial\hat{\Theta}/\partial\eta = \Pr\, \Theta_*$, and $\partial\hat{q}/\partial\eta = Sc_q\, u_*$. Thus, linear profiles have to be expected. The same is true when the respective eddy diffusivities become invariant with height. Such height invariance might be possible when the quiet regime prevails and the magnitude of the turbulent fluctuations is small across the entire ASL. Thus, we have to assume that Monin-Obukhov similarity is incomplete under strongly stable conditions. If under such conditions the constant flux approximation is no longer valid as debated, for instance, by Webb (1970) and Poulos and Burns (2003), Monin-Obukhov similarity must not be expected.

It is apparently that such physical and numerical shortcomings limit the accuracy of predictions of near-surface values of temperature and humidity by numerical models of the atmosphere when strongly stable stratification prevails. Mölders and Kramm (2009), for instance, used the Weather Research and Forecasting (WRF) model to simulate a five day cold weather period with multi-day inversions over Interior Alaska, where two different physical packages were applied alternatively. A Comparison of the simulations with radiosonde data and near-surface observations shows that WRF's performance for these inversions strongly depends on the physical packages chosen. Mölders and Kramm (2009) found, for instance, that the predicted near-surface air temperatures as well as the near-surface dew-point temperatures differ from the observations by up to 4 K, on average, depending on the physical packages used. All simulations had difficulties in capturing the full strength of the surface temperature inversion and in simulating strong variations of dew-point temperature profiles. The greatest discrepancies between simulated and observed vertical profiles of temperature and dew-point temperature occur around the levels of great wind shear.

Instead of the Businger-Dyer-Pandolfo relationship for momentum under unstable stratification, the conventional O'KEYPS formula[3],

$$\Phi_m^{\,4}(\zeta) - \gamma_{10}\,\zeta\,\Phi_m^{\,3}(\zeta) = 1 \qquad \text{for} \quad \zeta \le 0 \quad , \tag{2.35}$$

may alternatively be applied. It indicates a $\Phi_m(\zeta) \cong (-\gamma_{10}\,\zeta)^{-1/3}$ behavior for large negative Obukhov numbers, for which $\Phi_m(\zeta) << -\gamma_{10}\,\zeta$ becomes valid. The O'KEYPS formula with $\gamma_{10} = 9$ is experimentally proved for the range $-2 \le \zeta < 0$ (e.g., Businger et al., 1971); Panofsky and Dutton (1984), however, recommended a value of $\gamma_{10} = 15$. From a physical point of view the O'KEYPS formula seems to be more preferable than the Businger-Dyer-Pandolfo relationship because the former can be related to the local balance equation of the TKE. For horizontally homogeneous and steady-state conditions the non-dimensional form of this TKE equation reads

$$0 = -\,\Phi_d(\zeta) + \Phi_m(\zeta) - \zeta - \Phi_\varepsilon(\zeta) \quad , \tag{2.36}$$

where $\Phi_d(\zeta) = \Phi_E(\zeta) + \Phi_P(\zeta)$ represent the non-dimensional divergence of both the eddy flux of TKE, $E = \frac{1}{2}\,\overline{\rho\, w''\,\mathbf{v}''^2}$, denoted by $\Phi_E(\zeta)$ and the eddy flux $P \cong \overline{\rho\, w''\, p'}/\overline{\rho}$ resulting from pressure and vertical wind speed fluctuations and expressed by $\Phi_P(\zeta)$. Furthermore, $\Phi_\varepsilon(\zeta)$ is the similarity function of the energy dissipation $\hat{\varepsilon} = \overline{\varepsilon^*}/\overline{\rho}$. Relating the latter to the Heisenberg-von Weizsäcker law, $\hat{\varepsilon} = K_m^{\,3}\,\Lambda^{-4}$ (Heisenberg, 1948; von Weizsäcker, 1948) and postulating a mixing length for non-neutral conditions by $\Lambda = \Lambda_P \, \Phi_\Lambda(\zeta)$, with which Fortak (1969), Herbert and Panhans (1979), and Panhans and Herbert (1979) introduced the further similarity function $\Phi_\Lambda(\zeta)$ for improving the treatment of this length scale in dependence on non-neutral conditions, lead to

$$\Phi_m^{\,4}(\zeta) - \left(\frac{\Phi_d(\zeta)}{\zeta} - 1\right) \zeta \, \Phi_m^{\,3}(\zeta) = \Phi_\Lambda^{\,-4}(\zeta) \quad . \tag{2.37}$$

[3] O'KEYPS stands for the initials of various authors who proposed this formula (Obukhov, 1946; Kazansky and Monin, 1956; Ellison, 1957; Yamamoto, 1959; Panofsky, 1961; Sellers, 1962).

Herbert and Panhans (1979) and Panhans and Herbert (1979) also examined different expressions for $\Phi_\Lambda(\zeta)$. They found that its definition at the cost of an analytical hypothesis for the TKE-transport term leads to the most satisfactory agreement with the observational data by Wyngaard and Coté (1971).

The simplest possible case of interest, however, is a Prandtl-type mixing length for neutral stratification so that $\Lambda = \Lambda_P = u * \kappa \left(z - d\right)$. The omission of non-neutral effects in $\Phi_\Lambda(\zeta)$ supposes the argument that buoyancy and mean wind shear may generate turbulence in which the deviation of $\Phi_\Lambda(\zeta)$ from unity is too small to contribute significantly to the energy dissipation. This concept is usually employed in one-and-a-half-order closure schemes (e.g., Stull, 1988; Garratt, 1994). With this simplification, Eq. (2.37) becomes an extended version of the O'KEYPS formula given by (Kramm et al., 1996c)

$$\Phi_m^{\,4}(\zeta) - \left(\frac{\Phi_d(\zeta)}{\zeta} - 1\right) \zeta \, \Phi_m^{\,3}(\zeta) = 1 \quad . \tag{2.38}$$

Comparing this equation with formula (2.35) yields then

$$\gamma_{10} = \frac{\Phi_d(\zeta)}{\zeta} - 1 \quad , \tag{2.39}$$

i.e., it is unlikely that the quantity $\gamma_{10}$ is a constant, as already pointed out by Fortak (1969), Herbert and Panhans (1979), Panhans and Herbert (1979), and Kramm et al. (1996c). In contrast to the conventional O'KEYPS formula, the extended version (2.38) is not restricted to unstable stratification. The same is true in the case of formula (2.37).

Local similarity functions of the form

$$\Phi_m(\zeta) = \left(1 - \gamma_{11} \, \zeta\right)^{-\frac{1}{3}} \tag{2.40}$$

as found, for instance, by Carl et al. (1973) and Gavrilov and Petrov (1981) for unstable stratification in the range of $-10 \leq \zeta < 0$, reflect the same asymptotic behavior like the conventional O'KEYPS formula, but they disagree with that of the Businger-Dyer-Pandolfo relationship. Here, $\gamma_{11} = 15$ is assumed.

Following, for instance, Panofsky and Dutton (1984), Pal Arya (1988), as well as Kraus and Businger (1994) the formulae (2.24) and (2.25) should be used for practical purposes, where a von Kármán constant of $\kappa = 0.40$ has to be preferred. This recommendation together with (2.40) will be scrutinized in the 4th section. The effects of other values of the von Kármán constant as found, for instance, by Frenzen and Vogel (1995) and Andreas et al. (2006) are thoroughly discussed by Kramm and Herbert (2009).

## 3. Prandtl-Obukhov-Priestley scaling

Under free-convective conditions the Obukhov stability length is no longer relevant for the vertical profiles of mean values of wind speed, potential temperature, and specific humidity because the vertical transfer of momentum, sensible heat, and matter is rather independent of the friction velocity $u_*$ (Lumley and Panofsky, 1964). On the basis of the Prandtl-Obukhov-Priestley scaling one can derive the so-called $-4/3$ power law (e.g., Garratt, 1994; Kramm and Herbert, 2009)

$$\frac{\partial \widehat{\Theta}}{\partial z} = C \left( \frac{H}{c_{p,0}\, \bar{\rho}} \right)^{\frac{2}{3}} \left( \frac{g}{\Theta_m} \right)^{-\frac{1}{3}} \left( z - d \right)^{-\frac{4}{3}} \quad , \tag{3.1}$$

where $C \cong -1.07$ is the Priestley constant. Rearranging this equation in the sense of Monin-Obukhov scaling, where only dry air is considered (i.e., the influence of water vapor is ignored), leads to (e.g., Garratt, 1994; Kramm and Herbert, 2009)

$$\frac{\kappa \left( z - d \right)}{\Theta_*} \frac{\partial \widehat{\Theta}}{\partial z} = -C\, \kappa^{\frac{4}{3}} \left( -\zeta \right)^{-\frac{1}{3}} = 0.32 \left( -\zeta \right)^{-\frac{1}{3}} = \left( -30.5\, \zeta \right)^{-\frac{1}{3}} \quad . \tag{3.2}$$

It is obvious that formula (5.12) with $\Phi_h\left(\zeta\right) = \Phi_m{}^2\left(\zeta\right) = \left(1 - \gamma_2\, \zeta\right)^{-1/2}$ does not converge to the asymptotic solution (3.2) when $\zeta \ll 0$. Furthermore, the O'KEYPS formula (2.36) and the local similarity function (2.40) on the one hand and formula (3.2) on the other hand suggest that under free-convective conditions the $1/3$ power law, and, hence, formula (2.40) should be valid for both momentum and sensible heat.

Integrating Eq. (3.1) over the height interval $\left[z_r, z_R\right]$ leads to the Priestley-Estoque relation (e.g., Estoque, 1973)

$$\begin{Bmatrix} H \\ Q \end{Bmatrix} = -\bar{\rho}\,\Gamma_h \begin{Bmatrix} c_{p,0}\left(\widehat{\Theta_R} - \widehat{\Theta_r}\right) \\ \widehat{q_R} - \widehat{q_r} \end{Bmatrix} \tag{3.3}$$

with

$$\Gamma_h = \frac{C_1}{3\left\{\left(z_r - d\right)^{-\frac{1}{3}} - \left(z_R - d\right)^{-\frac{1}{3}}\right\}}\left(-\frac{\frac{g}{\Theta_m}\left(\widehat{\Theta_R} - \widehat{\Theta_r}\right)}{3\left\{\left(z_r - d\right)^{-\frac{1}{3}} - \left(z_R - d\right)^{-\frac{1}{3}}\right\}}\right)^{\frac{1}{2}}, \tag{3.4}$$

where, as in the case of forced-convective conditions, $Sc_{t,i} \cong Pr_t$ is assumed to determine the vertical transfer of water vapor also. The constant $C_1 \cong 0.90$ can be derived from Priestley's constant. Note that this expression is strongly sensitive to the choice of $z_r$.

As under free-convective conditions Monin-Obukhov scaling fails, Estoque (1973) proposed to calculate the friction velocity in the same manner like the vertical eddy fluxes of sensible heat and water vapor. Thus, one obtains

$$u_*^{\,2} = \Gamma_u\left(\widehat{u_R} - \widehat{u_r}\right) \tag{3.5}$$

with $\Gamma_u = \Gamma_h$. Unfortunately, Estoque's postulate is not scrutinized.

## 4. Assessing the integral similarity functions

It is well known that gradients of horizontal wind speed, temperature and humidity cannot be measured because of the limited spatial resolution of available sensors. This means that the true flux-gradient relationships and, hence, the local similarity functions are unsuitable for estimating the eddy fluxes of momentum, sensible heat and water vapor. Consequently, it is indispensable to relate these eddy fluxes, at least, to finite differences of horizontal wind speed, temperature, and humidity. This can be performed by integrating the non-dimensional gradients over the layer under study, where the constant flux assumptions (or approximations) are considered. The results of such integrations are customarily denoted as (vertical) profile functions (see formulae (2.10) to (2.13)).

Results from direct measurements of eddy fluxes and corresponding vertical profiles of the mean values of wind speed, temperature, and humidity obtained from concurrent measurements can be used to derive local similarity functions. Note that for the purpose of evaluation of such local similarity functions, quite independent data sets of directly measured eddy fluxes and mean vertical profiles even obtained concurrently are required. Data sets from field campaigns not considered for deriving such local similarity functions clearly satisfy this requirement (Kramm and Herbert, 2009).

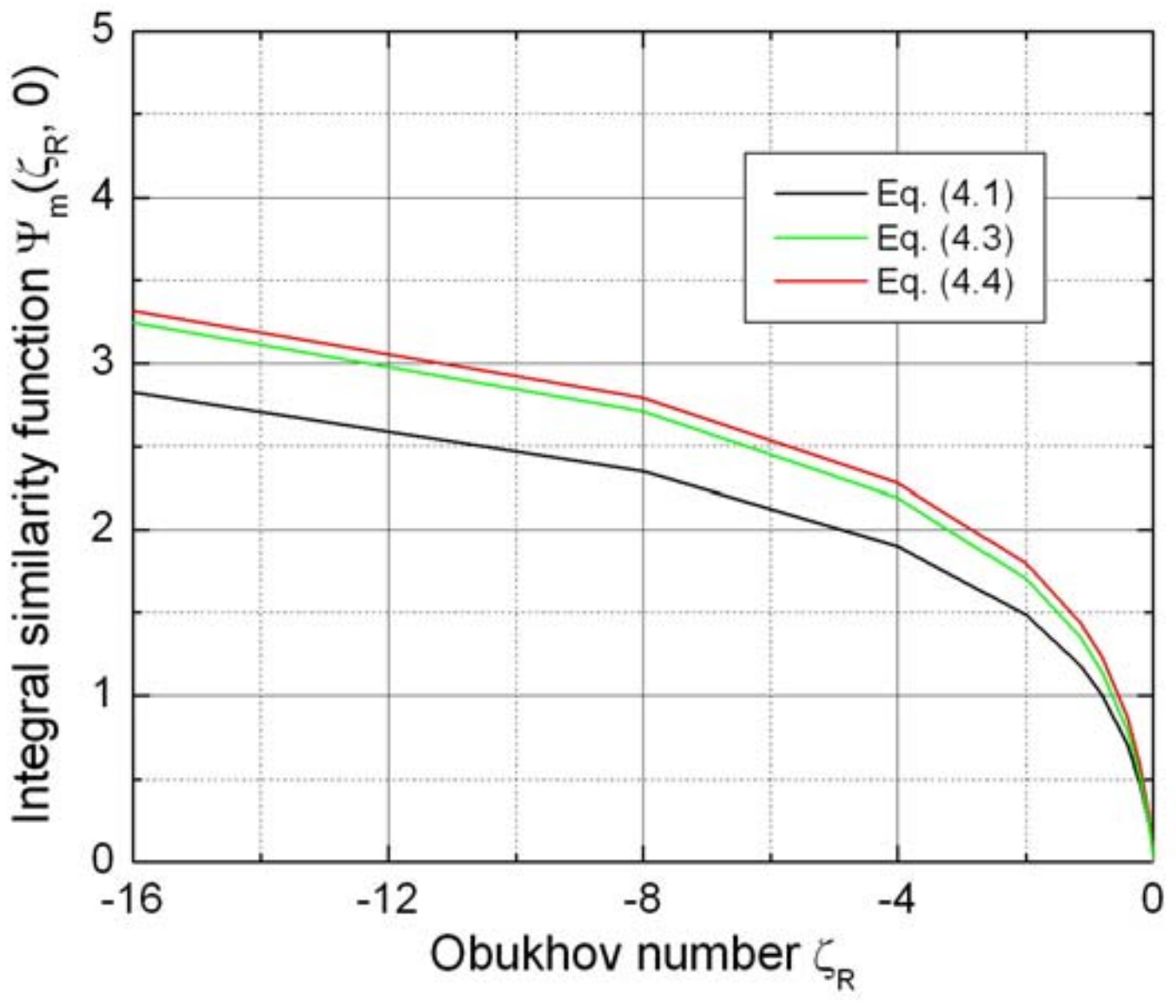

**Figure 10:** The integral similarity function $\Psi_m\left(\zeta_R, 0\right)$ for momentum obtained from formulae (4.1), (4.3), and (4.4) and plotted against the Obukhov number $\zeta_R$ (adopted from Kramm and Herbert, 2009).

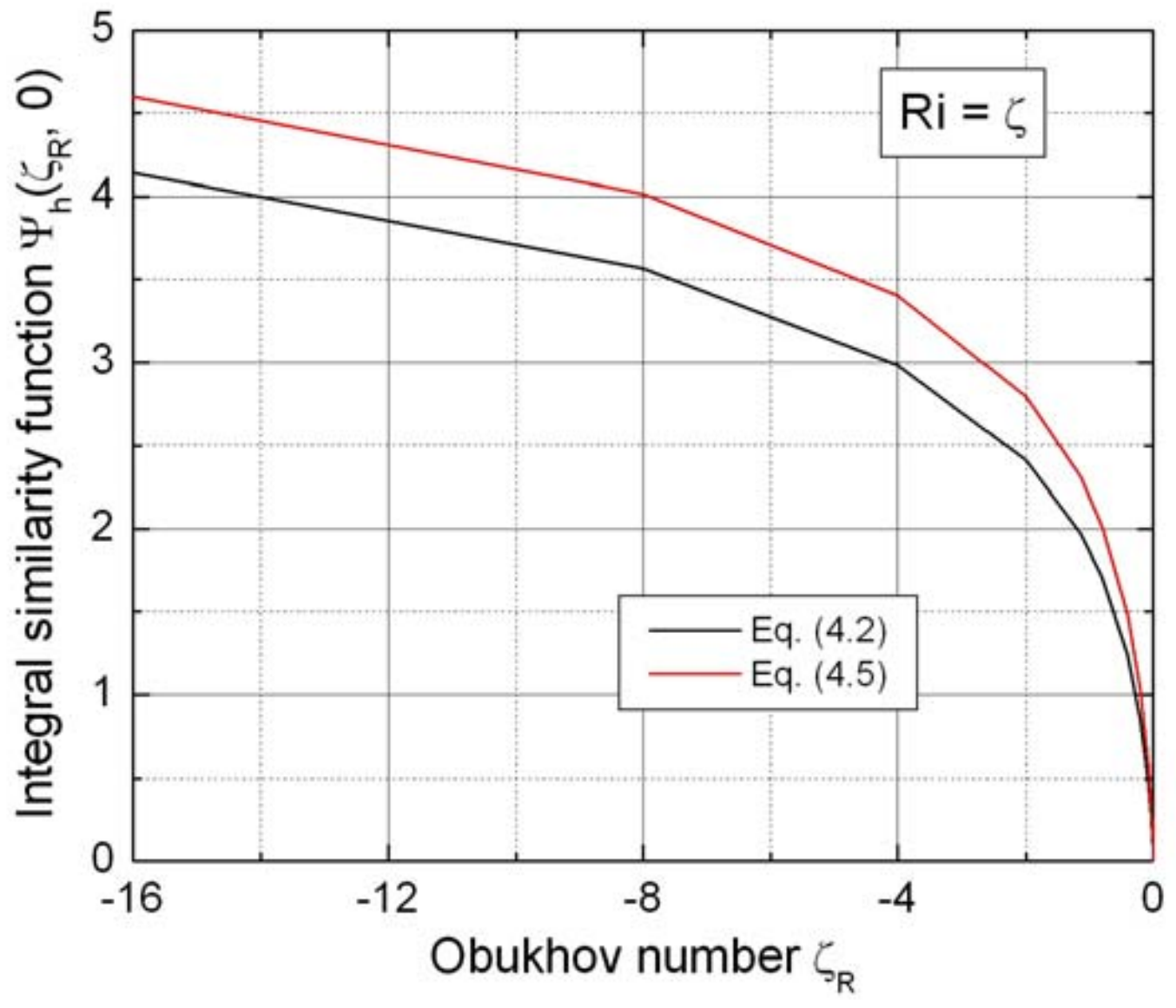

**Figure 11:** The integral similarity function $\Psi_h(\zeta_R, 0)$ for sensible heat obtained from formulae (4.2) and (4.5) and plotted against the Obukhov number $\zeta_R$ (adopted from Kramm and Herbert, 2009).

### 4.1 *Profile relations and integral similarity functions*

Inserting formulae (2.24) and (2.25) into the definition (2.13) yields (Kramm and Herbert, 1984; Kramm, 1989)

$$\Psi_m\left(\zeta_R, \zeta_r\right) = \begin{cases} -\gamma_1\left(\zeta_R - \zeta_r\right) & \text{for } L > 0 \\ 0 & \text{for } L \to \infty \\ 2\ln\dfrac{1+y_R}{1+y_r} + \ln\dfrac{1+y_R^{\,2}}{1+y_r^{\,2}} - 2\arctan\dfrac{y_R - y_r}{1+y_R\, y_r} & \text{for } L < 0 \end{cases} \tag{4.1}$$

and

$$\Psi_q\left(\zeta_R, \zeta_r\right) = \Psi_h\left(\zeta_R, \zeta_r\right) = \begin{cases} \Psi_m\left(\zeta_R, \zeta_r\right) & \text{for } L > 0 \\ 0 & \text{for } L \to \infty \\ 2\ln\dfrac{1+y_R^{\,2}}{1+y_r^{\,2}} & \text{for } L < 0 \end{cases} \tag{4.2}$$

with $y_{r,R} = \Phi_m^{\;-1}\left(\zeta_{r,R}\right) = \left(1 - \gamma_2\,\zeta_{r,R}\right)^{1/4}$, the reciprocal expressions of the local similarity functions in the unstable case at the two heights $z_r$ and $z_R$. Paulson's (1970) solutions substantially agree with formulae (4.1) and (4.2) when $y_r$ approaches to unity while $\zeta_r \to 0$. Inserting the conventional O'KEYPS formula (2.35) into definition (2.13) provides (Kramm and Herbert, 2009)

$$\left.\begin{aligned} \Psi_m\left(\zeta_R,\ \zeta_r\right) = \Phi_m\left(\zeta_r\right) - \Phi_m\left(\zeta_R\right) + 2\ln\frac{1+\Phi_m\left(\zeta_R\right)}{1+\Phi_m\left(\zeta_r\right)} + \ln\frac{1+\Phi_m^{\,2}\left(\zeta_R\right)}{1+\Phi_m^{\,2}\left(\zeta_r\right)} \\ + 2\ \arctan\frac{\Phi_m\left(\zeta_R\right) - \Phi_m\left(\zeta_r\right)}{1+\Phi_m\left(\zeta_R\right)\Phi_m\left(\zeta_r\right)} - 3\ln\frac{\Phi_m\left(\zeta_R\right)}{\Phi_m\left(\zeta_r\right)} \end{aligned}\right\} \quad \text{for } L \le 0 \tag{4.3}$$

with $\Phi_m\left(\zeta_{r,R}\right)$ provided by the conventional O'KEYPS formula. Obviously, the O'KEYPS solution (4.3) is more bulky than that obtained with the Businger-Dyer-Pandolfo relationship. This might be the reason why the latter is more widely used, even though the former has a stronger physical background. For $\zeta_r \to 0$, we have $\Phi_m(\zeta_r) \to 1$, and, hence, formula (4.3) approaches to Paulson's (1970) O'KEYPS-solution. If we consider formula (2.40), the definition (2.13) provides (Kramm and herbert, 2009)

$$\Psi_m\left(\zeta_R, \zeta_r\right) = \frac{3}{2} \ln \frac{y_R^{\ 2} + y_R + 1}{y_r^{\ 2} + y_r + 1} - \sqrt{3} \arctan \frac{x_R - x_r}{1 + x_R\ x_r} \quad \text{for } L \leq 0 \tag{4.4}$$

with $y_{r,R} = \Phi_m^{\ -1}\left(\zeta_{r,R}\right) = \left(1 - \gamma_{11}\ \zeta_{r,R}\right)^{1/3}$, the reciprocal expressions of the local similarity functions in the unstable case at the two heights $z_r$ and $z_R$, and $x_{r,R} = \left(2\ y_{r,R} + 1\right)/\sqrt{3}$. It approaches to Lettau's (1979) solution when $\zeta_r \to 0$.

Formulae (4.1), (4.3), and (4.4) are illustrated in Figure 10. As expected, formulae (4.3) and (4.4) only differ hardly when $\zeta$ tends to Obukhov numbers much smaller than zero which represent free-convective conditions. Simultaneously, the difference between Eq. (4.1) and the other two formulae grows continuously.

If we assume that $\Phi_h\left(\zeta\right) = \Phi_m^{\ 2}\left(\zeta\right) \Leftrightarrow Ri = \zeta$ and $\Phi_q\left(\zeta\right) = \Phi_h$ hold for the entire range of unstable stratification and that the local similarity function for momentum is given by Eq. (2.40), one obtains (Kramm and Herbert, 2009)

$$\Psi_q\left(\zeta_R, \zeta_r\right) = \Psi_h\left(\zeta_R, \zeta_r\right) = \frac{3}{2} \ln \frac{y_R^{\ 2} + y_R + 1}{y_r^{\ 2} + y_r + 1} + \sqrt{3} \arctan \frac{x_R - x_r}{1 + x_R\ x_r} \quad \text{for } L \leq 0 \quad . \tag{4.5}$$

Formulae (4.2) and (4.5) are illustrated in Figure 11. As shown, Eq. (4.5) provides appreciably larger values of $\Psi_h\left(\zeta_R, 0\right)$ than formula (4.2).

**Table 1:** Various combinations of integral similarity functions defined by Eq. (2.13) used in the profile functions (2.10) to (2.12).

| Model number | Equation number | | |
|---|---|---|---|
| | $\Psi_m$ | $\Psi_h$ | $\Psi_q$ |
| 1 | (4.1) | (4.2) | (4.2) |
| 2 | (4.4) | (4.5) | (4.5) |
| 3 | (4.4) | (4.2) | (4.2) |
| 4 | (4.4) | (4.4) | (4.4) |

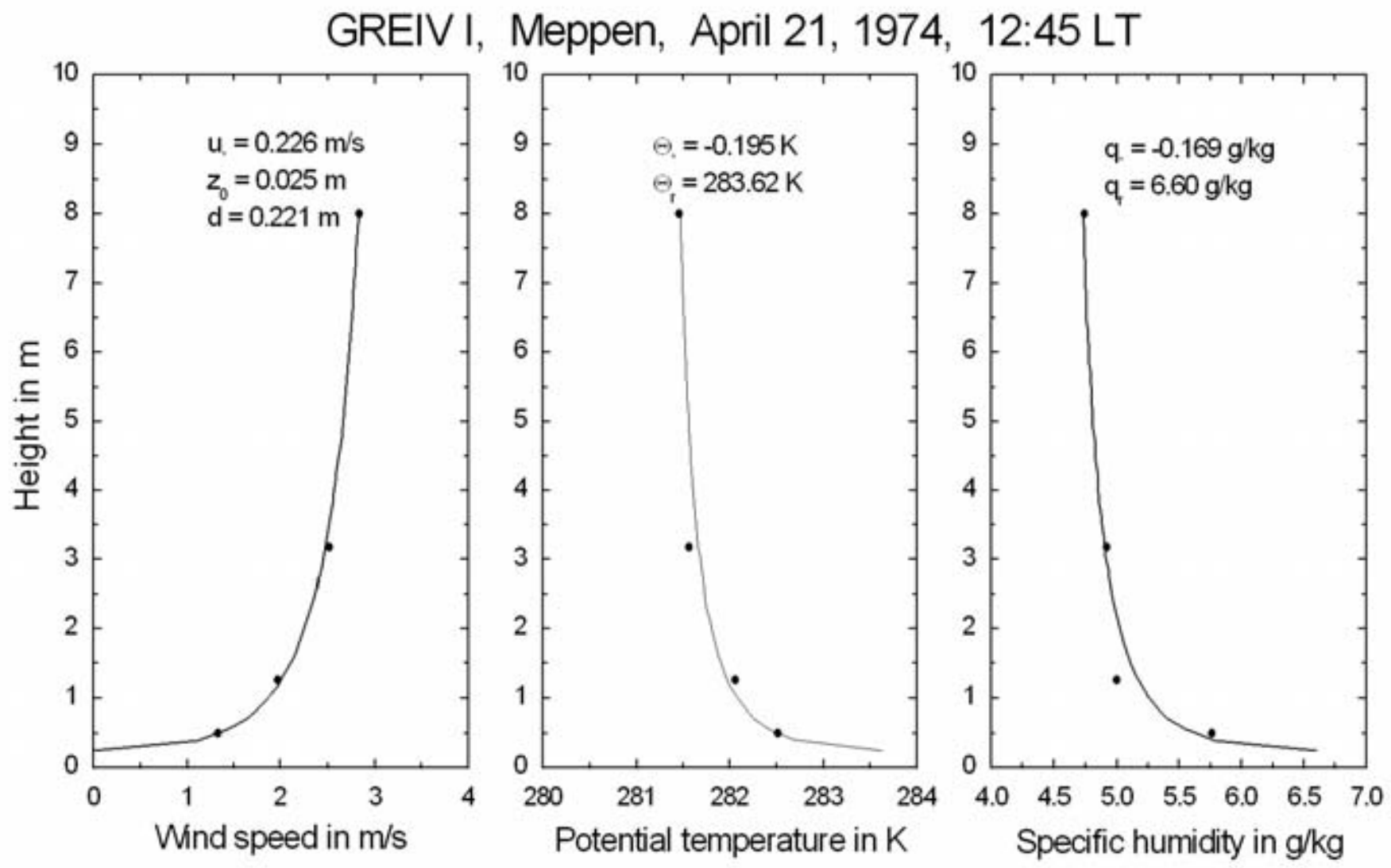

**Figure 12:** Typical vertical profile of wind speed, potential temperature and specific humidity predicted with Model 1 (see Table 1) for unstable stratification. The dots represent the observed values and the solid lines the calculated profiles (adopted from Kramm and Herbert, 2009).

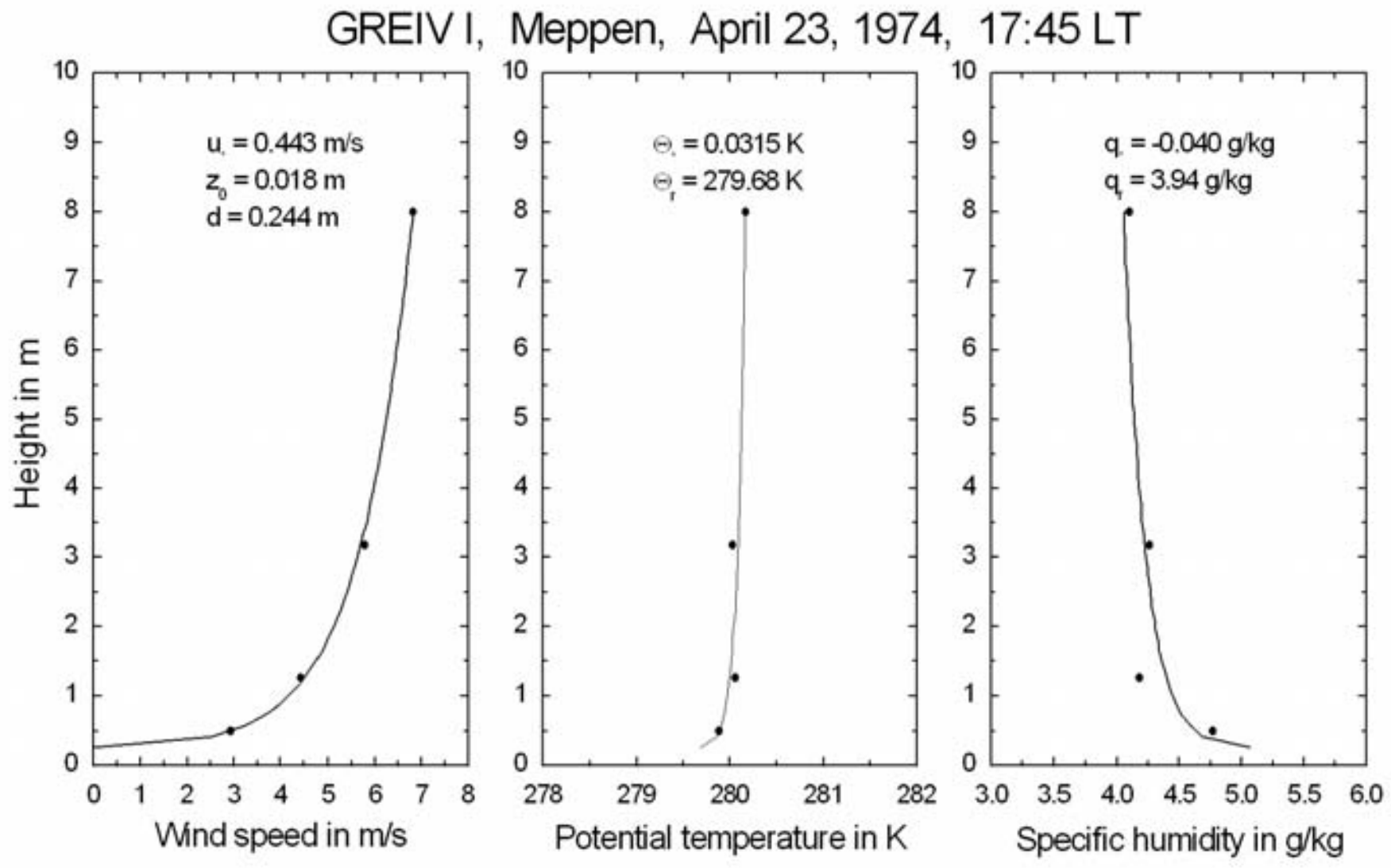

**Figure 13:** As in Figure 12, but for stable stratification.

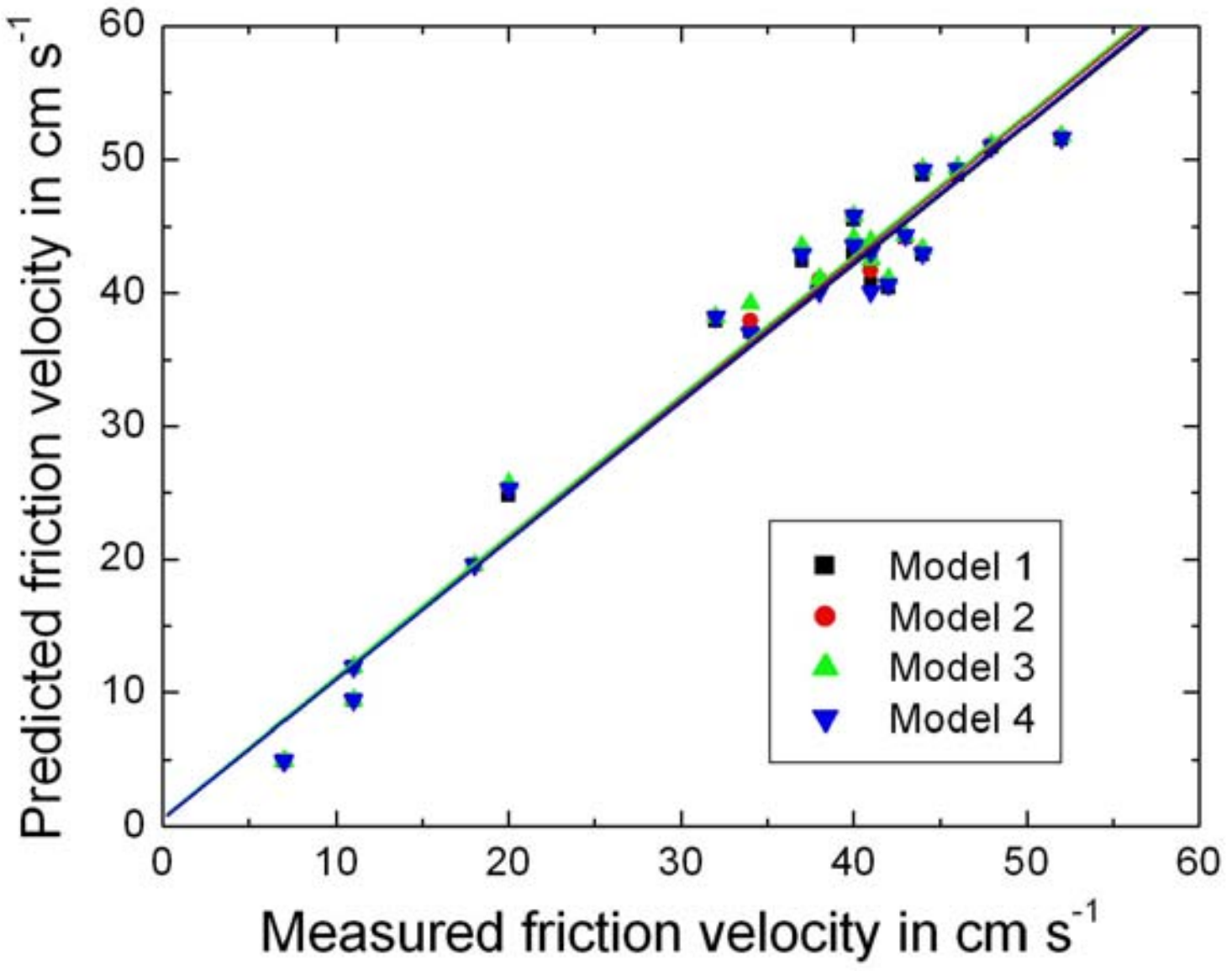

**Figure 14:** Predicted eddy fluxes of momentum (characterized by the friction velocity, $u_*$) vs. directly measured ones for four different parametric models. (data adopted from Kramm and Herbert, 2009). The colors of the lines are related to the colors of the symbols.

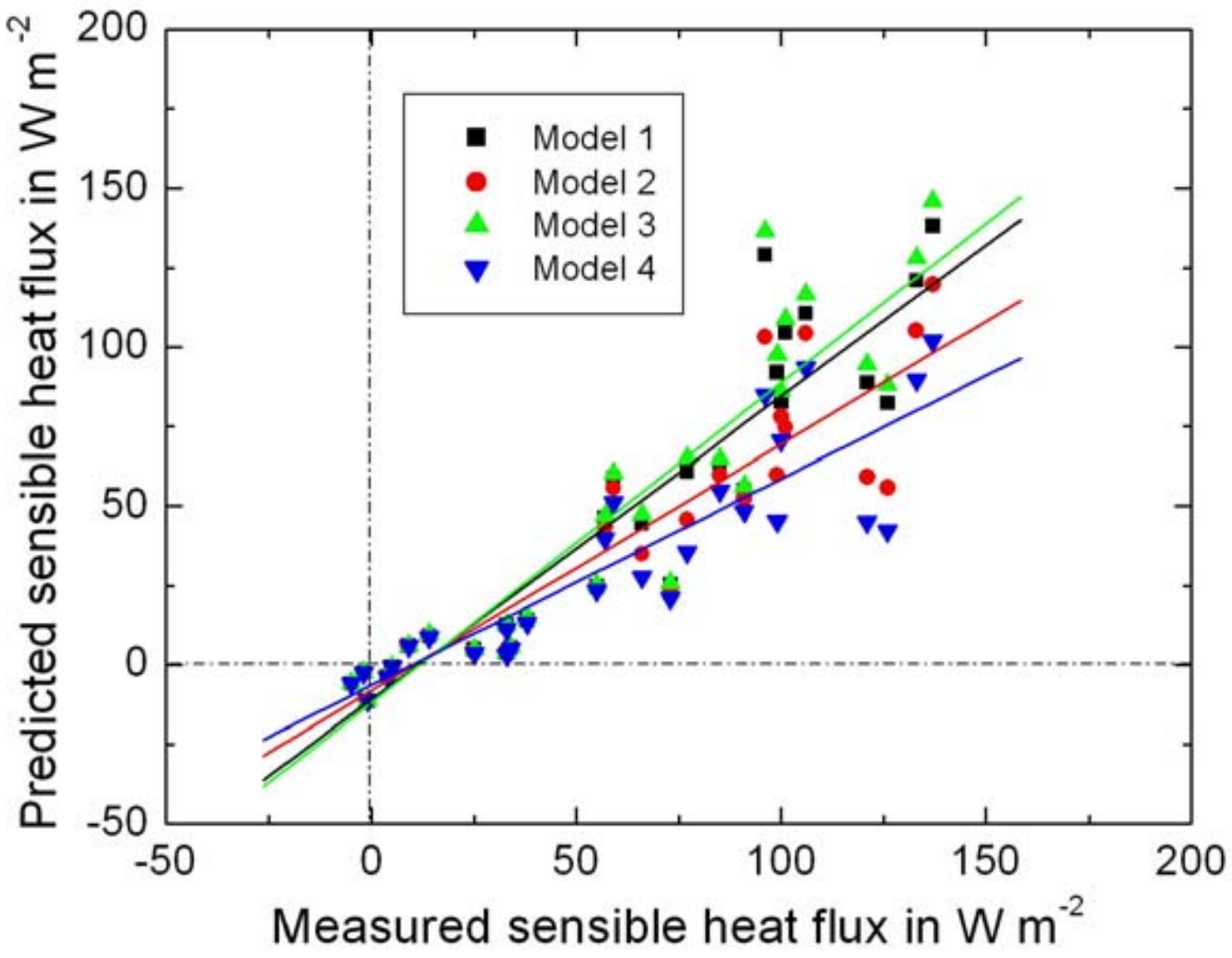

**Figure 15:** As in Figure 14, but for the eddy fluxes of sensible heat.

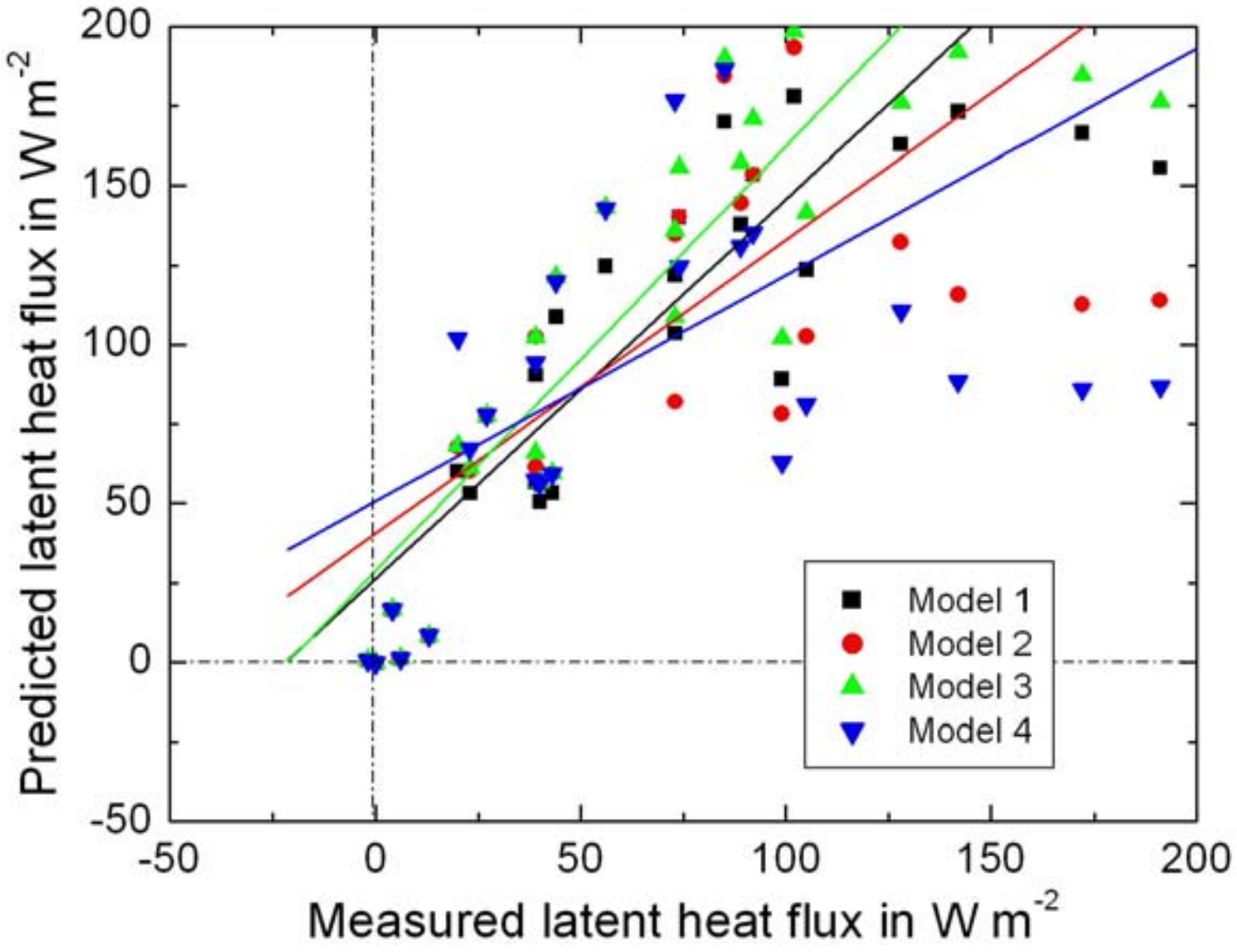

**Figure 16:** As in Figure 14, but for the fluxes of latent heat.

### *4.2 Computed eddy fluxes versus measured eddy fluxes*

As mentioned before, the equation sets (2.10) to (2.13) may be used to determine the scaling quantities $u_*$, $\Theta_*$, and $q_*$, and, hence the corresponding fluxes $\tau$, $H$, and $Q$, as well as the roughness length, $z_0$ (for $z_r = z_0 + d$), and the zero-plane displacement, $d$, from vertical profile measurements of wind speed, temperature and humidity (e.g., Stearns, 1970; Lo, 1977; Nieuwstadt, 1978; Kramm and Herbert, 1984; Kramm, 1989; Kramm et al., 1996c).

Results derived with the iteration procedure developed by Kramm and Herbert (1984, 2009) are illustrated in Figures 12 to 16. Figures 12 and 13 show examples of vertical profile of wind speed, potential temperature and specific humidity obtained from observed data collected during the GREIV I 1974 experiment. This experiment took place over a flat site covered with winter barley (about 0.25 m high) and rape (0.50 to 0.75 m high), near Meppen/Emsland in northern Germany in April 1974. Data sets of wind speed, dry- and wet-bulb temperatures (simultaneously measured 30-min averages) were obtained by groups from the Universities of Kiel (April 20 - 24, 1974) and Munich (April 24 - 27, 1974). The observations of the Kiel group were performed at heights of 0.5, 1.26, 3.18 and 8 m and those of the Munich group at heights of 0.5, 1, 2, 4, 8 and 16 m above ground. Both groups used Lambrecht cup anemometers and Frankenberger-type psychrometers. In addition, the 30-min run data of friction velocity and the vertical eddy fluxes of sensible and latent heat directly determined by the University of Mainz group using ultrasonic anemometer-thermometer (Kaijo-Denki 3D) and Lyman alpha hygrometer (self-developed) measurements were used for comparison. These fast-response measurements of the Mainz group were carried out in the vicinity of the instrumented mast of the Kiel group at a height of 2 m above ground. Note that the GREIV I 1974 data, fully documented by Beyer and Roth (1976), has not been used in deriving the universal functions on which the integral similarity functions presented here are based.

If $z_0 + d > z_1$ or more than 40 iteration steps had been required to determine $z_0$ and $d$, the profile data sets were generally rejected by the computer program. As mentioned before, such criteria occurred for profile data collected under very stable conditions with low wind speeds and temperature inversions or in the transition phase between lapse and inversion conditions, if stationary states required by the constant flux concept must not be expected (Stearns, 1970; Kramm, 1989).

Of 109 profile data sets of the Kiel group, 77 data sets were suitable for computation. From the 110 data sets of the Munich group, 73 an 69 data sets, respectively, were appropriate for computation, based on vertical profiles which included 5 and 6 levels (with and without the 16 m level of observation). The two instances of different thermal stratification illustrated in Figures 12 and 13 show that the calculated least squares fits coincide very well with the values observed.

Generally, compared with the predicted eddy fluxes of sensible and latent heat the predicted eddy fluxes of momentum characterized by $u_*$ much better agree with the directly measured fluxes for all four models (see Figure 14). As illustrated in Figures 15 and 16 the predicted eddy fluxes of sensible and latent heat show a large scatter independent of the model used. Note, however, that the sampling intervals of the vertical profile measurements and the direct measurements of eddy fluxes differ by about 15 min.

In the case of Model 1 general agreement also exist between the friction velocity, and the eddy fluxes of sensible and latent heat predicted with Model 1 (see Table 1) and those directly measured (see Figures 14 to 16). It is called the reference case because the combination of integral similarity functions as gathered in Model 1 is usually recommended (e.g., Panofsky and Dutton, 1984; Pal Arya, 1988; Kraus and Businger, 1994).

As illustrated in Figure 10, the results provided by formulae (4.3) and (4.4) only differ hardly when $\zeta$ tends to large negative Obukhov numbers that represent free-convective conditions. Since formula (4.3) is more bulky than formula (4.4), it seems to be reasonable to replace Eq. (4.1) that does not match free-convective conditions by formula (4.4) and, with respect to $\Phi_h(\zeta) = \Phi_m{}^2(\zeta) \Leftrightarrow Ri = \zeta$, Eq. (4.2) by formula (4.5) when unstable stratification of air is considered (see Model 2 in Table 1). Compared with the reference case, this combination of formulae provides eddy flux results that more disagree with those directly determined (see Figures 14 to 16). In comparison with the reference case, a slightly better agreement especially for the eddy flux of sensible heat can be achieved when for unstable stratification only Eq. (4.1) is replaced by formula (4.4) (see Model 3 in Table 1). Consequently, $Ri = \zeta$ is not longer valid. As illustrated in Figure 17, this combination leads to $|Ri| \geq |\zeta|$. The eddy flux results obtained with the Model 3 are illustrated in Figures 14 to 16, too.

As mentioned before, the local similarity function (2.40) as found, for instance, by Carl et al. (1973) as well as Gavrilov and Petrov (1981) for unstable stratification and the conventional O'KEYPS formula (2.35) on the one hand and Eq. (3.2) on the other hand suggest that under free-convective conditions the one-thirds law, and, hence, Eq. (4.4) should be valid for both momentum and sensible heat (and water vapor) as Estoque (1973) postulated by $\Gamma_u = \Gamma_h$ (see section 3). However, using Model 4 that is based on formula (4.4) yields rather insufficient results for the vertical eddy fluxes of sensible and latent heat (see Figures 14 to 16, Model 4). Consequently, Estoque's (1973) postulate has to be considered with care.

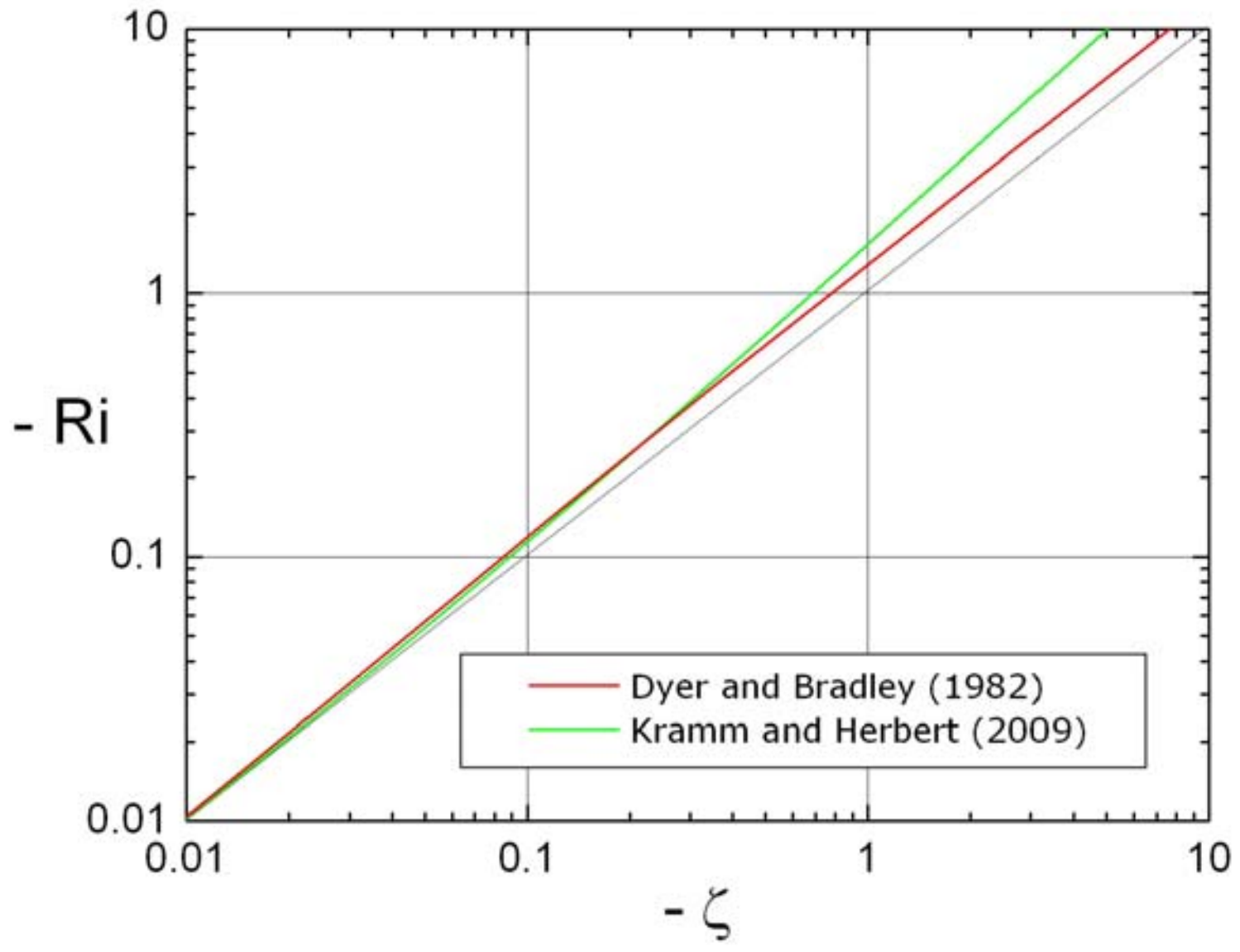

**Figure 17:** Gradient Richardson number, $Ri$, versus Obukhov number, $\zeta$. The one-to-one line represents $\Phi_h(\zeta) = \Phi_m{}^2(\zeta)$ that leads to $Ri = \zeta$ (see formula (2.27)).

### *4.3 The impact of the von Kármán constant*

In all calculations a von Kármán constant of $\kappa = 0.4$ was chosen. This value differs from that of Businger et al. (1971), Frenzen and Vogel (1995), and Andreas et al. (2006). Even though the value of the von Kármán constant, $\kappa = 0.387 \pm 0.003$, derived by Andreas et al. (2006) for near-neutral stratification, is based on the largest, most comprehensive atmospheric data set ever used, this value has to be confirmed for wide ranges of non-neutral stratification.

Nevertheless, to assess the impact of a value for the von Kármán constant not simultaneously derived with the local similarity functions of momentum, sensible heat and matter, we consider principles of Gaussian error propagation (GEP). The deviation, for instance, of the friction velocity owing to the deviation of the von Kármán constant from its original value can be expressed by (e.g. Kreyszig, 1970; Kramm and Herbert, 2009)

$$\delta u_* = \pm \frac{\partial u_*}{\partial \kappa} \delta \kappa \tag{4.6}$$

with

$$\frac{\partial u_*}{\partial \kappa} = \frac{u_*}{\kappa}\left(1 + \frac{\kappa}{\ln \dfrac{z_R - d}{z_r - d} - \Psi_m(\zeta_R, \zeta_r)} \frac{\partial \Psi_m(\zeta_R, \zeta_r)}{\partial \kappa}\right) \quad . \tag{4.7}$$

The derivative $\partial \Psi_m(\zeta_R, \zeta_r)/\partial \kappa$ depends on the integral similarity function used (see subsection 4.1). For neutral conditions we simply obtain $\partial u_*/\partial \kappa = u_*/\kappa$ and in a further step for the relative deviation of the friction velocity

$$\frac{\delta u_*}{u_*} = \pm \frac{\delta \kappa}{\kappa} \quad . \tag{4.8}$$

Thus, if $\kappa = 0.35$ is the value simultaneously derived with the local similarity functions for momentum and sensible heat by Businger et al. (1971), but the "true" value of $\kappa = 0.387$ is used, the relative deviation of $u_*$ will amount to $10.6\,\%$. For the purpose of comparison: As the sum of the fluxes of sensible and latent heat corresponds to nearly $H + E \approx 100\ \mathrm{W/m^2}$, on average, the uncertainty that can be related to the von Kármán constant is appreciably larger than the globally averaged net anthropogenic radiative forcing in 2005 that correspond to $RF = 1.6\,(0.6\ \text{to}\ 2.4)\,\mathrm{W\,m^{-2}}$ relative to pre-industrial conditions defined at 1750 (Forster et al., 2007). Even the small uncertainty in the von Kármán constant of $\Delta\kappa = \pm\,0.003$ is large enough to provide an uncertainty in this result of $H + E \approx 100\ \mathrm{W/m^2}$ which is similar to the mean value of the globally averaged net anthropogenic radiative forcing mentioned before.

### *4.4 The prediction of uncertainties*

In the subsection before we discussed the impact of the von Kármán constant on the flux results because of (a) an inappropriate choice of its value that amounts to a procedural error and (b) its stochastic error with which any empirical quantity is fraught. The former has to be removed or, at least, minimized, but the impact by the latter has to be addressed by predicting the inherent uncertainty during a model simulation for assessing the reliability of the predicted results. This was done, for instance, by Milford et al. (1995) for chemical mechanisms with the framework of urban and regional scale oxidant modeling, and by Mölders et al (2005) for the simulation of soil processes.

To determine the model uncertainty caused by $n$ empirical parameters, $\chi_j$, $i = j, \ldots, n$, for $m$ quantities, $\phi_i$, $i = 1, \ldots, m$, like the fluxes of sensible and latent heat at the Earth's surface, one may consider GEP principles (e.g., Mölders et al., 2005). In doing so, the equation for predicting a quantity $\phi_i$ is derived for all empirical parameters $\chi_i$ on which it depends. The standard deviation (statistical uncertainty) of the predicted quantity can be calculated from these individual derivations $\partial\phi_i / \partial\chi_j$ and the standard deviations $\sigma_{\chi_j}$ of the j[th] empirical parameters $\chi_j$ by (e.g. Kreyszig, 1970)

$$\delta\phi_i = \pm\sqrt{\sum_{j=1}^{n}\left(\frac{\partial\phi_i}{\partial\chi_j}\right)^2 \delta^2\chi_j} = \pm\sqrt{\sum_{j=1}^{n}\left\{\phi_i, \delta\chi_j\right\}^2} \quad \text{for } i = 1, \ldots, m \quad , \tag{4.9}$$

where $\delta^2\chi_j$ are the variances;

$$\frac{\partial\phi_i}{\partial\chi_j}\,\delta\chi_j = \left\{\phi_i,\, \delta\chi_j\right\} \tag{4.10}$$

and $\delta\phi_i$ are denoted contribution (term) and uncertainty, respectively. The relative error is defined by $\varepsilon_{\phi_i} = \delta\phi_i / \phi_i$ (see, e.g. Eq. (4.8)).

## 5. Flux aggregation

Natural surfaces of a landscape are often heterogeneous over virtually all scales, i.e., also over the resolvable scales considered in weather and general circulation models (e.g., Mölders and Raabe, 1996; Giorgi and Avissar, 1997; Kramm et al., 2004). Thus, the exchange momentum, of sensible heat, and water vapor (and long-lived trace constituents like carbon dioxide ($CO_2$)) at the interface land surface-atmosphere of any grid element of such models cannot be treated by assuming homogeneous or dominant land-use types. To address this heterogeneity of a landscape in numerical models of the atmosphere it is indispensable to aggregate the fluxes mentioned before over the patchy fields of any grid element.

Different strategies have been developed to consider subgrid-scale surface heterogeneity of the patchy surface of a grid element, for instance, by averaging surface properties (e.g., Lhomme, 1992; Dolman, 1992) or by *statistic-dynamic* approaches (e.g., Wetzel and Chang, 1988; Entekhabi and Eagleson, 1989; Avissar, 1992). Computationally more expensive concepts to consider the properties of a heterogeneous surface are flux-related strategies like the *mosaic approach* (Avissar and Pielke, 1989), the conventional *blending-height concept* (e.g., Wieringa, 1986; Mason, 1988; Claussen, 1991; 1995), the *explicit-subgrid scheme* (Seth et al., 1994; Mölders et al., 1996), the *vertically extended explicit-subgrid scheme* (Tetzlaff et al., 2002), or the *mixture approach*, wherein for the different surface types tightly coupled energy balance equations are solved (e.g., Deardorff, 1978; Sellers et al., 1986; Kramm et al., 1996a).

Recently, Kramm et al. (2007) presented and discussed an improved version of the *blending-height concept*. In contrast to its conventional form, this version of the blending height concept is based on a consistent formulation of the *mosaic approach* by including the thermal stratification of air in the ASL. Since the prediction of the blending height is related to thermal stratification of air in the ASL, it can be varied during the diurnal cycle.

## 5.1 *The conventional blending-height concept*

The conventional blending-height concept can be considered as a mosaic approach for momentum. This mosaic approach is related to (Avissar and Pielke, 1989)

$$F_j^k = \sum_{i=1}^{N} \alpha_{i,j} \, F_{i,j}^{\;k} \tag{5.1}$$

Here, $F_j^k$ is the representative of a flux of the j[th] grid element, where $k = 1$ stands for momentum, $k = 2$ for sensible heat, and $k = 3$ for water vapor. Furthermore, $\alpha_{i,j}$ is the fractional area of the j[th] grid element covered by the i[th] patch. The quantity $F_j^k$ is considered as an area-weighted one, where $F_{i,j}^{\;k}$ is the corresponding flux provided by the i[th] patch with a given land-use type from the N patches that compose the j[th] grid cell. Note that in the following the subscripts have always the same meaning. The mosaic approach was introduced into the literature by Avissar and Pielke (1989). Mölders et al. (1996) evaluated it by comparison with the *explicit-subgrid strategy* of Seth et al. (1994). As illustrated by Mölders et al. (1996), the mosaic approach gathers all patches of a grid element having the same land-use type by ignoring their true location within that grid element.

In accord with Eq. (5.1) we may write for $k = 1$

$$F_{i,j}^{\;1} = \frac{\overline{\rho_j} \, \kappa^2 \left\{ \widehat{u_j}\left(h_{b,j}\right) \right\}^2}{\left\{ \ln\left( \dfrac{h_{b,j}}{z_{0,i,j}} \right) \right\}^2} \quad . \tag{5.2}$$

This equation is based on the assumption that a logarithmic wind profile is prevailing over the i[th] patch of the j[th] grid cell with the roughness length $z_{0,i,j}$. This means that (a) only neutrally stratified air is considered, and (b) the mean horizontal wind speed at the height $z_{0,i,j}$ is always equal to zero. The latter, of course, is not entirely true. Even in the neutrally stratified ASL the mean horizontal wind speed only extrapolates to zero when the height above ground is approaching to $z_{0,i,j}$ (or $z_{0,i,j} + d_{i,j}$, when a zero-plane displacement, $d_{i,j}$, is introduced additionally). This mean wind speed is not exactly equal to zero, but it may be ignored because it is usually much smaller than the mean wind speed in the region of the ASL aloft, as postulated, for instance, by the logarithmic wind profile always associated with a positive wind shear.

The magnitude of the aggregated flux of momentum, $F_j^1$, is then given by

$$F_j^1 = \overline{\rho}_j\, \kappa^2 \left\{ \widehat{u}_j \left( h_{b,j} \right) \right\}^2 \sum_{i=1}^{N} \frac{\alpha_{i,j}}{\left\{ \ln \left( \frac{h_{b,j}}{z_{0,i,j}} \right) \right\}^2} \quad , \tag{5.3}$$

Assuming that $F_{i,j}^{\,1}$ can also be determined in the same manner like any $F_j^1$ leads then to the relation

$$\frac{1}{\left\{ \ln \left( \frac{h_{b,j}}{z_{0,j}} \right) \right\}^2} = \sum_{i=1}^{N} \frac{\alpha_{i,j}}{\left\{ \ln \left( \frac{h_{b,j}}{z_{0,i,j}} \right) \right\}^2}$$

that may be used to define an aggregated roughness length, $z_{0,j}$, for the $j^{th}$ grid cell. The key quantity in this equation is the blending height $h_{b,j}$ for which the mean horizontal wind speed, $\widehat{u}\left(h_{b,j}\right)$, should be approximately in equilibrium with the local surface and also independent of horizontal position (e.g., Mason, 1988; Claussen, 1991, 1995). This blending height is usually unknown. Therefore, some formulae were proposed by several authors, to estimate it. Mason's (1988), for instance, heuristically proposed,

$$\frac{h_{b,j}}{L_{c,j}} \left\{ \ln \left( \frac{h_{b,j}}{z_{0,j}} \right) \right\}^2 \cong 2\, \kappa^2 \tag{5.4}$$

Whereas Claussen (1991) suggested

$$\frac{h_{b,j}}{L_{c,j}} \left\{ \ln \left( \frac{h_{b,j}}{z_{0,j}} \right) \right\}^2 \cong C_b\, \kappa \quad . \tag{5.5}$$

Here $C_b \cong 1.75$ is an empirical constant, and $L_{c,j}$ is the horizontal scale of aerodynamic roughness variation. These formulae also demand the existence of a logarithmic wind profile.

Obviously, $F_j^1$ and $z_{0,j}$ are related to the level of the blending height. From this point of view the blending-height concept may be acceptable to accommodate the atmospheric flow above $h_{b,j}$ to the heterogeneous distribution of surface properties in a given grid cell like surface roughness. As proposed, for instance, by von Salzen et al. (1996), the aggregated momentum flux, and in a further step, the aggregated friction velocity, related to the aggregated momentum flux by $u_{*,j}^{\;2} = F_j^1 \big/ \overline{\rho}_j$, should further be used to calculate the bulk transfer coefficients for sensible heat and water vapor similar to formulae (2.18) and (2.19).

*5.2 Effects owing to horizontal heterogeneity*

As aforementioned, horizontal homogeneity of the micrometeorological field quantities like the mean values of horizontal wind speed, temperature, and humidity is a major prerequisite for micrometeorological scaling according to either Monin and Obukhov (1954) for forced-convective conditions or Prandt (1932), Obukhov(1946), and Priestley(1959) for free-convective conditions. As already demonstrated by Bernhardt (1990), in the case of patchy surfaces horizontal homogeneity of these micrometeorological field quantities cannot be expected (Kramm et al., 2004). By assuming steady-state conditions, neutral stratification and well-established logarithmic wind profiles over each of the patches of different aerodynamic roughness and by considering the equation of continuity for the x-z-plane, Bernhardt (1990) obtained for the mean vertical wind speed

$$\widehat{w}(x,z) = -\frac{1}{\kappa}\left\{z\,\frac{\partial u_*(x)}{\partial x}\ln\left(\frac{z}{z_0(x)}\right) - (z - z_0)\left\{\frac{\partial u_*(x)}{\partial x} + \frac{u_*(x)}{z_0(x)}\frac{\partial z_0(x)}{\partial x}\right\}\right\} \quad , \tag{5.6}$$

i.e., even in the case of neutral stratification, a change in surface roughness can produce a logarithmic-linear vertical profile of mean vertical wind speed as well as of friction stress (Bernhardt, 1990).

The conditions of the constant flux approximation, on which the blending-height concept and, hence, the subsequent aggregation of the fluxes of sensible heat and water vapor (and long-lived trace species) is based, may not be fulfilled when advective effects become too large so that the blending-height concept can produce results of insufficient degrees of accuracy. Therefore, it is indispensable to estimate such advective effects. This can simply be performed by considering the equation of continuity for the x-z-plane, where an incompressible medium (i.e., density effects are ignored) is assumed for which $\nabla \cdot \overline{\mathbf{v}} = 0$ is an acceptable approximation of the equation of continuity. Here, $\overline{\mathbf{v}}$ is the vector of the mean wind speed, where $\overline{u}$, $\overline{v}$, and $\overline{w}$ are its components with respect to the horizontal co-ordinates, $x$ and $y$, and the vertical co-ordinate, $z$, respectively. Thus, for the x-z-plane, for instance, we have

$$\frac{\partial \overline{u}}{\partial x} + \frac{\partial \overline{w}}{\partial z} = 0 \tag{5.7}$$

Assuming logarithmic wind profiles for neutral stratification of air, as requested by the conventional blending-height concept, the variation of the horizontal wind speed over an area of the $j^{th}$ grid element comprising two adjacent local patches of roughness lengths $z_{0,1,j}$ and $z_{0,2,j}$ can be expressed by

$$\frac{\partial \bar{u}(x,z)}{\partial x} \cong \frac{\Delta \bar{u}(x,z)}{\Delta x} = \frac{\bar{u}(x_{2,j},z) - \bar{u}(x_{1,j},z)}{L_{c,j}} = \frac{\bar{u}(h_{b,j})}{L_{c,j}} \left\{ \frac{\ln \frac{z}{z_{0,2,j}}}{\ln \frac{h_{b,j}}{z_{0,2}}} - \frac{\ln \frac{z}{z_{0,1,j}}}{\ln \frac{h_{b,j}}{z_{0,1}}} \right\} , \tag{5.8}$$

where $L_{c,j}$ is the horizontal scale of the roughness variation. The mean vertical velocity component, $\bar{w}(x,z)$, can be calculated using Eq. (5.2). One obtains (e.g., Panofsky, 1973; Bernhardt, 1990)

$$\frac{\Delta \bar{u}(x,z)}{\Delta x} \cong -\frac{\Delta \bar{w}(x,z)}{\Delta z} = -\frac{\bar{w}(x,z)}{z} \tag{5.9}$$

or

$$\bar{w}(x,z) = -\bar{u}(h_{b,j}) \frac{z}{L_{c,j}} \left\{ \frac{\ln \frac{z}{z_{0,2,j}}}{\ln \frac{h_{b,j}}{z_{0,2,j}}} - \frac{\ln \frac{z}{z_{0,1,j}}}{\ln \frac{h_{b,j}}{z_{0,1,j}}} \right\} . \tag{5.10}$$

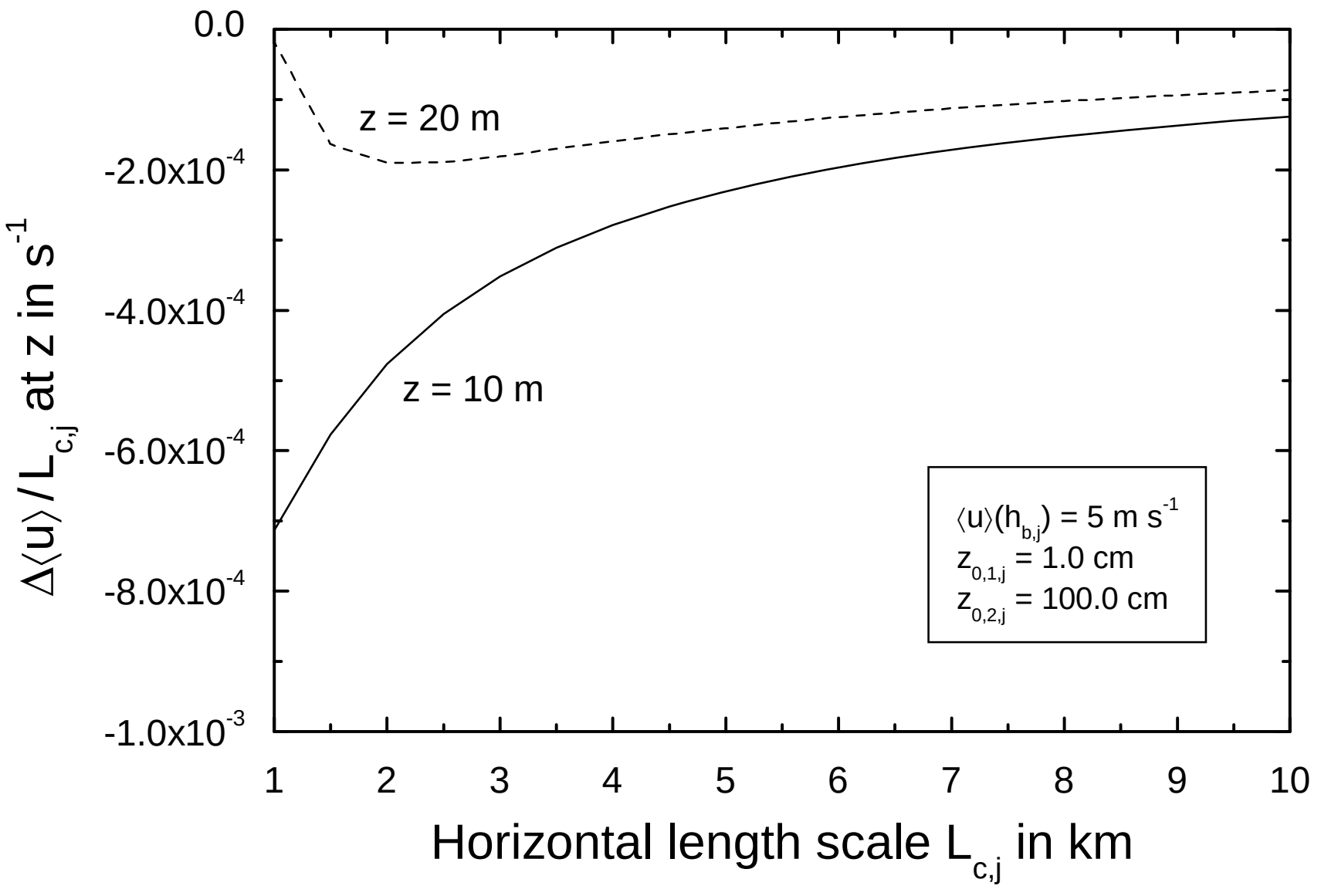

**Figure 18:** Variation of the mean horizontal wind speed, $\left\langle \mathrm{u} \right\rangle = \overline{\mathrm{u}}$, versus horizontal length scale, $\mathrm{L}_{\mathrm{c,j}}$, for two different levels below the blending height, $\mathrm{h}_{\mathrm{b,j}}$ (adopted from Kramm et al., 2007).

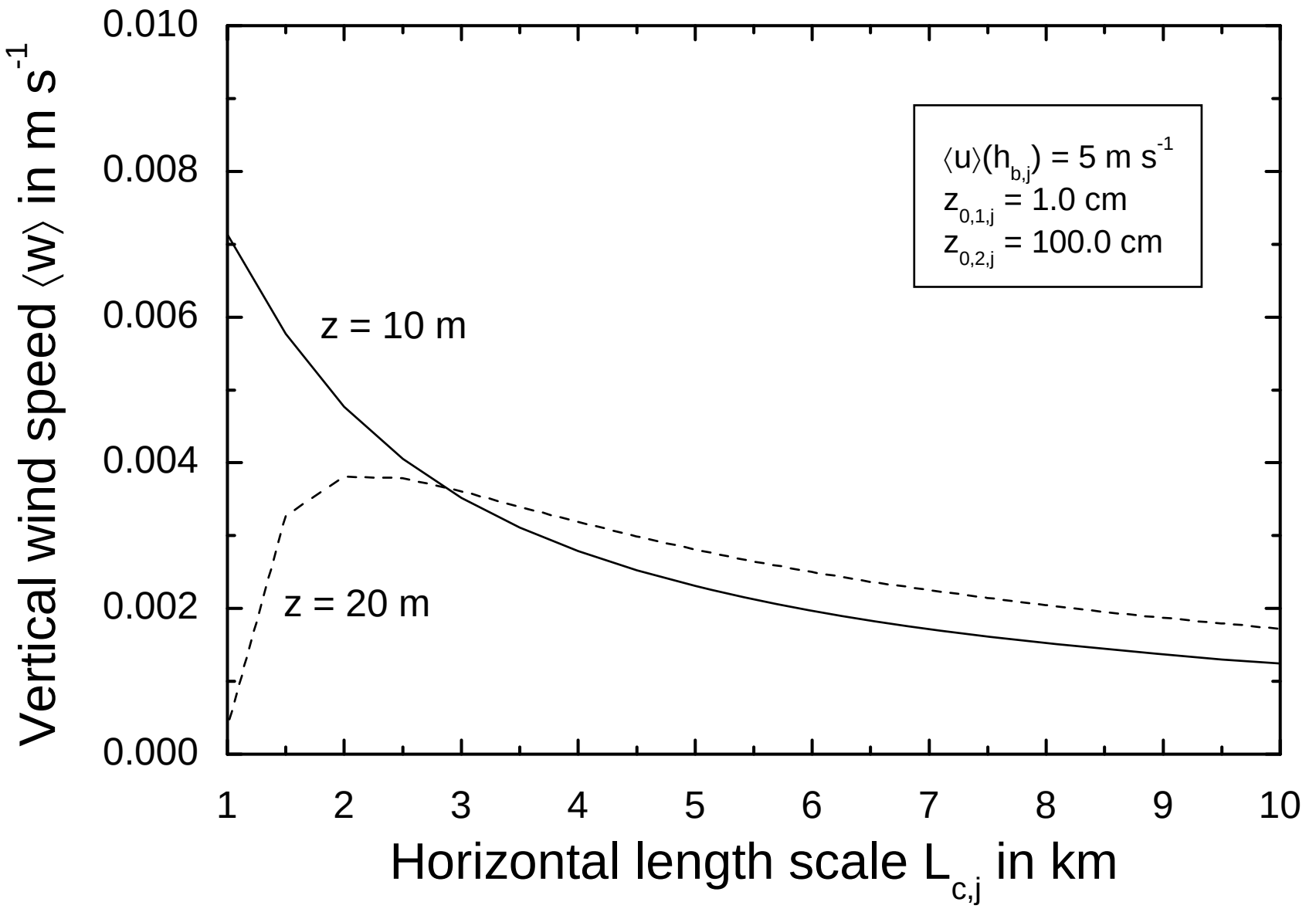

**Figure 19:** Calculated mean vertical wind speed, $\left\langle \mathrm{w} \right\rangle = \overline{\mathrm{w}}$, versus horizontal length scale, $\mathrm{L}_{\mathrm{c,j}}$, for two different levels below the blending height, $\mathrm{h}_{\mathrm{b,j}}$ (adopted from Kramm et al., 2007).

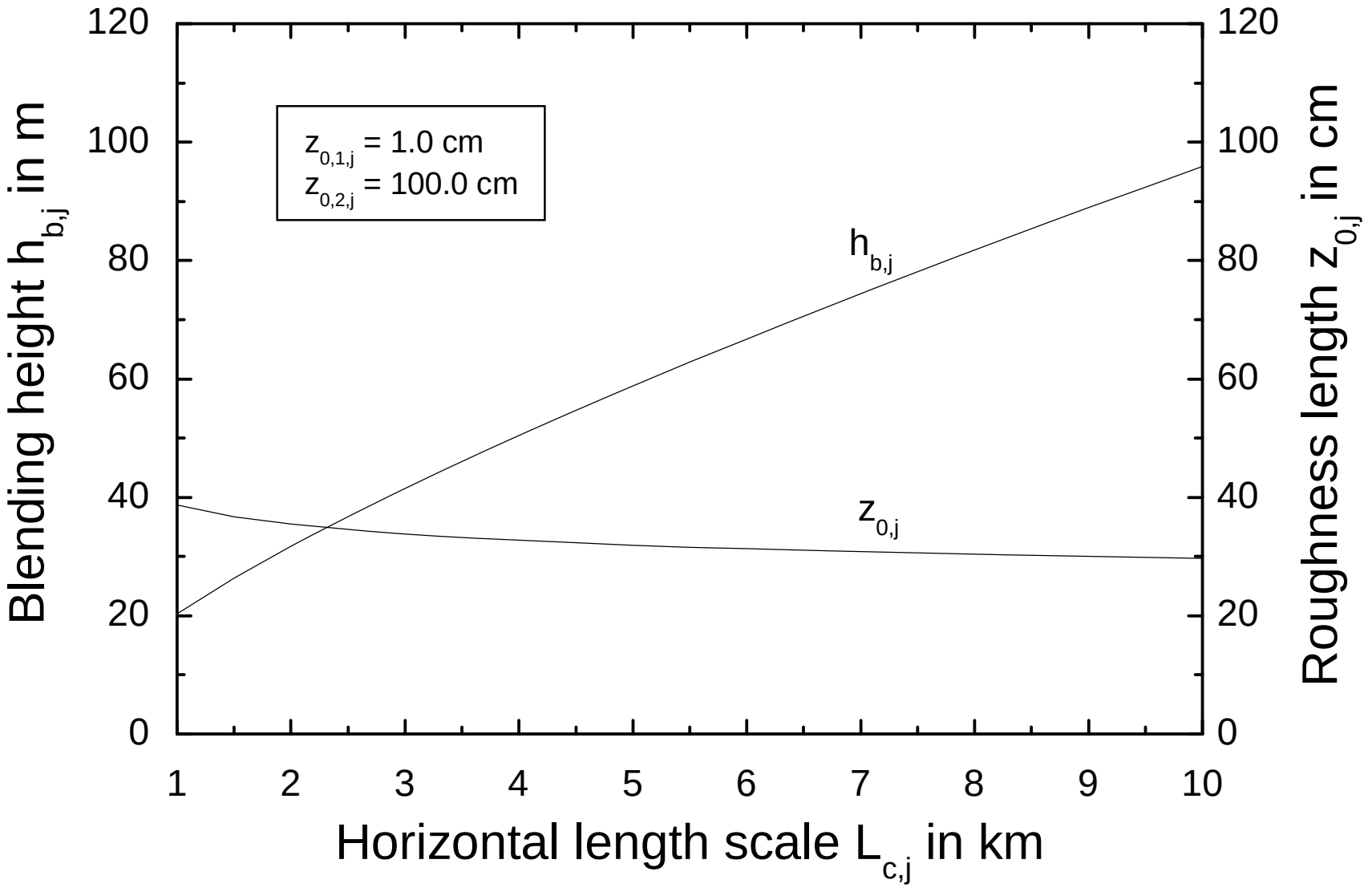

**Figure 20:** Blending height, $h_{b,j}$, and aggregated roughness length, $z_{0,a}$, versus horizontal length scale, $L_{c,j}$. The relationship between $h_{b,j}$ and $L_{c,j}$ was calculated with Mason's (1988) heuristic formula (5.4).

Assuming, for instance, $z_{0,1,j} = 1.0\ \mathrm{cm}$ (e.g., grassland), $z_{0,2,j} = 100.0\ \mathrm{cm}$ (e.g., forest), $h_{b,j} = 70.0\ \mathrm{m}$, and $z = 10.0\ \mathrm{m}$ provides $\Delta\bar{u}\left(x,z\right)\big/\Delta x \approx -\,0.24\ \bar{u}\left(h_{b,j}\right)\big/L_{c,j}$. As shown in Figures 18 and 19, the horizontal wind speed appreciably varies over the distance $L_{c,j}$ which leads to appreciable values of $\bar{w}\left(x,z\right)$. These estimates agree with the results provided by a numerical model of the atmospheric boundary layer which does not include a constant flux layer approach (Estoque, 1973, and the references therein). Thus, especially the vertical transfer of sensible heat and water vapor close to the Earth's surface can strongly be affected by such a mean vertical motion. The calculated mean vertical wind speed, for instance, at the height $z = 10\ \mathrm{m}$ (see Figure 19) is for smaller values of $L_{c,j}$ of the same order of magnitude than the so-called exchange velocity defined by

$$v_{\chi,k} = \frac{\bar{F}_j^{\,k}}{\bar{\rho}\,\bar{\chi}_k} \quad , \qquad \text{for } k = 2,\, 3 \tag{5.11}$$

so that the criterion (Kramm et al., 2004)

$$\left|\bar{\rho}\,\bar{w}\,\bar{\chi}_k\right| << \left|v_{\chi,k}\,\bar{\rho}\,\bar{\chi}_k\right| \Leftrightarrow \left|\bar{w}\right| << \left|v_{\chi,k}\right| \tag{5.12}$$

is notably violated. Note that in the case of mass transfer the exchange velocity, $v_\chi$, is denoted as exhalation (emission) velocity when the flux of matter is positive (i.e., directed upward); negative values of the exchange velocity are, therefore, related to deposition processes. Consequently, the use of aggregated bulk transfer coefficients to calculate the fluxes of sensible heat and water vapor as proposed, for instance, by von Salzen et al. (1996), where a momentum flux aggregated by the blending-height concept, and in a further step, an aggregated friction velocity is taken into account, has to be assessed as inconsistent. In other words: The blending-height concept denies the existence of the equation of continuity and the consequences based on it. Note that in Figure 20 the calculated blending height and the aggregated roughness length $z_{0,j}$ are plotted versus $L_{c,j}$. To derive the $h_{b,j} - L_{c,j}$ relationship from the values of $z_{0,1,j}$ and $z_{0,2,j}$, Mason's (1988) heuristic approach (5.4) was applied.

The deviation from the horizontal homogeneity will become insignificant and the relationships for the constant flux layer presented above may sufficiently be suitable, when the criterion (5.12) is fulfilled. To guarantee that this criterion is fulfilled during field experiments, especially in the case of occasionally observed internal boundary layers, the level, $z_R$, above ground has to fulfill the fetch requirement, for instance, for neutral stratification (e.g., Raabe, 1983)

$$z_R \le z_{max} \approx 0.3\,\delta^{1/2} \quad , \tag{5.13}$$

where the fetch, $\delta$, is the distance between the change of the surface properties in the weather-site region of a measuring tower and that tower itself, and $z_{max}$ is the height of the undisturbed new equilibrium layer (Rao et al., 1974). If we assume, for instance, $z_{max}$ = 10 m, the fetch will amount to a practicable value of $\delta \approx 1.1$ km. Since reliable results of constant flux layer relationships (see subsection 2.2.3) were derived under the auspices of fulfilled fetch requirements, horizontal grid resolutions may not be finer than those suggested by condition (4.5) or similar criteria for stable and unstable stratification (see, e.g., Rao et al., 1974; Garratt, 1990) if the constant flux relationships presented above are to be applied.

## *5.3 An improved blending-height concept*

To guarantee that the blending-height concept is entirely self-consistent, it is indispensable (a) to include diabatic effects, (b) to calculate the corresponding area- weighted fluxes of sensible heat, water vapor, momentum, and long-lived trace constituents in accord with the principles of the mosaic approach defined by Eq. (5.1), and (c) to guarantee that the effects of the inferred mean vertical velocity, $\hat{w}$, can be ignored. The first two requirements can be satisfied by the following set of equations (Kramm et al., 2007; see also formulae (2.14) to (2.16)):

$$F_j^1 = \tau_j = \overline{\rho}_j \left\{ \widehat{u}_j\left(h_{b,j}\right) \right\}^2 \sum_{i=1}^{N} \alpha_{i,j}\, C_{m,i,j} \quad , \tag{5.14}$$

$$F_j^2 = H_j = -\, c_{p,d}\, \overline{\rho}_j\, \widehat{u}_j\left(h_{b,j}\right) \sum_{i=1}^{N} \alpha_{i,j}\, C_{h,i,j} \left\{ \widehat{\Theta}\left(h_{b,j}\right) - \widehat{T_{s,i,j}} \right\} \quad , \tag{5.15}$$

and

$$F_j^3 = Q_j = -\, \overline{\rho}_j\, \widehat{u}_j\left(h_{b,j}\right) \sum_{i=1}^{N} \alpha_{i,j}\, C_{q,i,j} \left\{ \widehat{q}\left(h_{b,j}\right) - \widehat{q_{s,i,j}} \right\} \tag{5.16}$$

where the drag coefficient and the transfer coefficients for sensible heat and water vapor are given by (see also formulae (2.17) to (2.19))

$$C_{m,i,j} = \frac{\kappa^2}{\left( \kappa \left( \frac{\xi_{d,i,j}}{2} \right)^{-\frac{1}{2}} + \ln \frac{h_{b,j}}{z_{r,i,j}} - \Psi_m\left(\zeta_{b,j}, \zeta_{r,i,j}\right) \right)^2} \quad , \tag{5.17}$$

$$C_{h,i,j} = \frac{\kappa^2}{\left( \kappa \left( \frac{\xi_{d,i,j}}{2} \right)^{-\frac{1}{2}} + \ln \frac{h_{b,j}}{z_{r,i,j}} - \Psi_m\left(\zeta_{b,j}, \zeta_{r,i,j}\right) \right) \left( \kappa\, B_{h,i,j}^{\ -1} + \ln \frac{h_{b,j}}{z_{r,i,j}} - \Psi_h\left(\zeta_{b,j}, \zeta_{r,i,j}\right) \right)} \quad , \tag{5.18}$$

and

$$C_{q,i,j} = \frac{\kappa^2}{\left(\kappa\left(\frac{\xi_{d.i,j}}{2}\right)^{-\frac{1}{2}} + \ln\frac{h_{b,j}}{z_{r,i,j}} - \Psi_m\left(\zeta_{b,j}, \zeta_{r,i,j}\right)\right)\left(\kappa\, B_{q,i,j}^{\ -1} + \ln\frac{h_{b,j}}{z_{r,i,j}} - \Psi_q\left(\zeta_{b,j}, \zeta_{r,i,j}\right)\right)} \quad . (5.19)$$

The upper and the lower boundaries of the integral similarity functions with respect to the i[th] patch,

$$\Psi_{m,h,q}\left(\zeta_{b,j}, \zeta_{r,i,j}\right) = \int_{\zeta_{r,i,j}}^{\zeta_{b,j}} \frac{1 - \Phi_{m,h,q}\left(\zeta\right)}{\zeta}\, d\zeta \qquad (5.20)$$

are given by $\zeta_{b,i,j} = h_{b,j} / L_{i,j}$ and $\zeta_{r,i,j} = z_{r,i,j} / L_{i,j}$, where $L_{i,j}$ is the corresponding Obukhov stability length related to the eddy fluxes of momentum, sensible heat and water vapor for this patch.

If we define

$$C_{\chi,j} = \sum_{i=1}^{N} \alpha_{i,j}\, C_{\chi,i,j} \qquad (5.21)$$

and

$$X_{s,j} = \frac{1}{C_{\chi,j}} \sum_{i=1}^{N} \alpha_{i,j}\, C_{\chi,i,j}\, X_{s,i,j} \quad , \qquad (5.22)$$

where $C_{\chi,j}$ in Eq. (5.21) stands for the aggregated drag coefficient ($\chi = m$) and the aggregated bulk transfer coefficients for sensible heat ($\chi = h$) and water vapor ($\chi = q$), respectively, and $X$ in formula (5.22) stands for temperature, $\widehat{\Theta}$, and humidity, $\hat{q}$, respectively, we may write

$$F_j^1 = \tau_j = \overline{\rho}_j\, C_{m,j} \left\{\widehat{u}_j\left(h_{b,j}\right)\right\}^2 \quad , \qquad (5.23)$$

$$F_j^2 = H_j = -\, c_{p,d}\, \overline{\rho}_j\, C_{h,j}\, \widehat{u}_j\left(h_{b,j}\right)\left(\widehat{\Theta}\left(h_{b,j}\right) - \widehat{T_{s,j}}\right) \quad , \qquad (5.24)$$

and

$$F_j^3 = Q_j = -\overline{\rho}_j\, C_{q,j}\, \widehat{u}_j\left(h_{b,j}\right)\left(\widehat{q}\left(h_{b,j}\right) - \widehat{q_{s,j}}\right) \quad . \tag{5.25}$$

Equation (5.22) substantiates that appropriate averaging the surface properties, as suggested by Lhomme (1992) and Dolman (1992), requires that $C_{\chi,i,j} = C_{\chi,j}$ is valid for all patches of a grid element so that $X_{s,j} = \sum_{i=1}^{N} \alpha_{i,j}\, X_{s,i,j}$ . It is unlikely that over a patchy surface of a grid element this requirement is generally fulfilled.

Analogous to formula (5.4) the blending height is related to the horizontal length scale by

$$\frac{h_{b,j}}{L_{c,j}} \left\{ \kappa \left( \frac{\xi_{d.j}}{2} \right)^{-\frac{1}{2}} + \ln \left( \frac{h_{b,j}}{z_{r,j}} \right) - \Psi_m \left( \zeta_{b,j}, \zeta_{r,j} \right) \right\}^2 \cong 2\,\kappa^2 \quad , \tag{5.26}$$

where $\xi_{d,j}$, $z_{r,j}$ and $\zeta_{r,j}$ should be interpreted as the aggregated ones.

To guarantee that the constant flux approximation is valid, the ratio $h_{b,j}/L_{c,j}$ has to be related to criterion (5.12).

## 6. Final remarks and conclusions

Since the prediction of climate is mainly considered as a prediction of second kind, it is indispensable to assess the accuracy with which these boundary conditions can be determined so that we can find a reasonable answer, whether climate is predictable with a sufficient degree of accuracy or not.

Thus, our contribution was mainly focused on the accuracy of the parameterization of the fluxes of sensible heat and water vapor required for solving the couple set of energy and water flux balance equations for the Earth's surface. The parameterization schemes for the interfacial sublayer in the immediate vicinity of the Earth's surface and for the fully turbulent layer above presented here document that an appreciable degree of uncertainty exists. The great uncertainty, inherent in the universal functions on which the integral similarity functions assessed before are based, is reflected by the considerable scatter in the results of sophisticated field campaigns. This uncertainty affects also the results for the gradient-Richardson number, the turbulent Prandtl number, $Pr_t$, and the turbulent Schmidt number, $Sc_{t,q}$, (and the turbulent Lewis-Semenov number, $LS_{t,q}$) for water vapor customarily used in such parameterization schemes. Especially for strongly stable stratification further research is urgently required because it seems that in this stability range Monin-Obukhov similarity becomes incomplete. Based on dimensional arguments, the Prandtl-Obukhov-Priestley similarity may be adequate for free-convective conditions, but its verification demands a sufficient degree of empirical evidence.

Even though the value of the von Kármán constant, $\kappa = 0.387 \pm 0.003$, derived by Andreas et al. (2006), is based on the largest, most comprehensive atmospheric data set ever used, this value has to be confirmed for wide ranges of non-neutral stratification. Since the local similarity functions and the von Kármán constant have to be determined simultaneously, the use of independently determined ones may produce an uncertainty that is larger than the globally averaged net anthropogenic radiative forcing in 2005 of $RF = 1.6\left(0.6 \text{ to } 2.4\right) W\ m^{-2}$ relative to pre-industrial conditions defined at 1750 (Forster et al., 2007). Even the small uncertainty in the von Kármán constant of $\Delta\kappa = \pm\ 0.003$ is large enough to provide an uncertainty in this result of $H + E \approx 100\ W/m^2$ which is similar to the value of RF.

In contrast to the O'KEYPS formula and the local similarity function for momentum of Carl et al. (1973) and Gavrilov and Petrov (1981), the Businger-Dyer-Pandolfo relationship does not converge to such free-convective conditions. Because of the Businger-Pandolfo relationship, $\Phi_h\left(\zeta\right) = \Phi_m{}^2\left(\zeta\right)$, the same is true in the case of the similarity function for sensible heat. Lumley and Panofsky (1964) already pointed out that the question of the relative size of $K_h$ and $K_m$, the eddy diffusivities for sensible heat and momentum, respectively, has still not been answered satisfactorily. We have to recognize that, even forty-five years later, their statement is further valid. Hitherto, values of the turbulent Prandtl number, the turbulent Schmidt number and the turbulent Lewis-Semenov number determined for the ASL are still scarce.

These results provided by the different parameterization schemes substantiate that a great uncertainty exists in the prediction of the eddy fluxes of sensible and latent heat. This degree of uncertainty may become still larger when the soil is covered by canopies of vegetation. In 1995 John Monteith stated: Describing the theory for an early type of porometer, Penman (1941) wrote that '*a compromise must be found between obtaining results that cannot be interpreted and obtaining no results that can be interpreted with precision*'. Monteith pointed out that *a similar dilemma is posed by attempts to model the accommodation of transpiring vegetation with the atmosphere. Separate, complex models are available for both regimes but a combined model would be an unmanageable monster from which useful output would be extremely hard to obtain. I have tried to combine much simpler models but am well aware that the predictions derived in this way lack precision and some may even prove to be wrong in detail.*

With respect to climate predictions especially for high latitude regions like the Arctic, the current degree of uncertainty seems to be too large. Thus, more direct eddy flux measurements are necessary for improving such parameterization schemes and, in a further step, for notably reducing the inherent uncertainty. The same is true in the case of flux aggregation principles indispensable for considering landscapes of patchy fields in general circulation models with coarse horizontal grid resolutions.

It is not surprising to us that the National Science Foundation (NSF) recently announced solicitation 09-568, Climate Process and Modeling Teams (CPT), where the key aim of the CPT concept is to speed development of global coupled climate models and reduce uncertainties in climate models by bringing together theoreticians, field observationalists, process modelers and the large modeling centers to concentrate on the scientific problems facing climate models today.

However, it is time to acknowledge that geophysical principles are burdened by notably uncertainties. This is not a tragedy, it is the reality. There is an urgent requirement to predict these uncertainties during long-term integrations to yield climate projections.